\documentclass[final,leqno,onefignum,onetabnum]{siamltex1213}

\usepackage{amssymb,amsmath}
\usepackage{color}
\usepackage[lined,boxruled]{algorithm2e}
\usepackage{comment}

\newtheorem{rem}{Remark}
\title{Updating and downdating techniques for optimizing network communicability}
\author{Francesca Arrigo\footnotemark[2] 
\and Michele Benzi\footnotemark[3]}
\begin{document}
\maketitle

\renewcommand{\thefootnote}{\fnsymbol{footnote}}
\footnotetext[2]{Department of Science and High Technology, 
University of Insubria, Como 22100, Italy (\email 
francesca.arrigo@uninsubria.it).}
\footnotetext[3]{Department of Mathematics and Computer science, 
Emory University, Atlanta, Georgia 30322, USA (\email benzi@mathcs.emory.edu).
The work of this author was supported by National Science Foundation grants
DMS1115692 and DMS-1418889.} 
\renewcommand{\thefootnote}{\arabic{footnote}}

%%%%%%%%%%%%%%%%%%%%%%%%%%%%%%%%%%%%%%%%%%%%%%%%%%%%%%%%%%%%%%%%%%%%%%%%%%
%%%%%%%%%%%%%%%%%%%%%%%%%% ABSTRACT %%%%%%%%%%%%%%%%%%%%%%%%%%%%%%%%%%%%%%
%%%%%%%%%%%%%%%%%%%%%%%%%%%%%%%%%%%%%%%%%%%%%%%%%%%%%%%%%%%%%%%%%%%%%%%%%%
\begin{abstract}
The total communicability of a network (or graph) is
defined as the sum of the entries in the exponential
of the adjacency matrix of the network, possibly
normalized by the number of nodes. This quantity
offers a good measure of how easily information spreads
across the network, and
can be useful in the design of networks having
certain desirable properties. The total communicability
can be computed quickly even for large networks
using techniques based on the Lanczos algorithm.

In this work we introduce some heuristics that can be
used to add, delete, or rewire a limited number of edges in a given
sparse network so that the modified network has a large
total communicability. %These
%methods are based on updating, downdating and rewiring
%techniques that take into account the change in total
%communicability resulting from the addition or deletion
%of an edge. 
To this end, we introduce new edge centrality measures
which can be used to guide in the selection of edges to be added or
removed.  

Moreover, we show experimentally that the total communicability 
provides an effective and easily computable measure of how ``well-connected"
a sparse network is. 

\end{abstract}

\begin{keywords}
network analysis; eigenvector centrality; subgraph centrality; total communicability; 
edge centrality; free energy; natural connectivity.
\end{keywords}

\begin{AMS}
05C82, 15A16, 65F60
\end{AMS}
\pagestyle{myheadings}
\thispagestyle{plain}
\markboth{{\sc Francesca Arrigo and Michele Benzi}}{{Network updating and downdating}}

%%%%%%%%%%%%%%%%%%%%%%%%%%%%%%%%%%%%%%%%%%%%%%%%%%%%%%%%%%%%%%%%%%%%%%%%%
%%%%%%%%%%%%%%%%%%%%%%%%% INTRODUCTION %%%%%%%%%%%%%%%%%%%%%%%%%%%%%%%%%%
%%%%%%%%%%%%%%%%%%%%%%%%%%%%%%%%%%%%%%%%%%%%%%%%%%%%%%%%%%%%%%%%%%%%%%%%%

\section{Introduction}
Network models are nowadays ubiquitous in the natural, information, social,
and engineering sciences. The last 15 years or so have seen the emergence of the
vast, multidisciplinary field of Network Science, with contributions
from a wide array of researchers including physicists, mathematicians, 
computer scientists, engineers, biologists, and social scientists 
\cite{Linked,EstradaBook,NewmanBook}. Applications of Network
Science range from biology to public health, from social network analysis to   
homeland security, from economics to the humanities, from marketing
to information retrieval. Network analysis is
also an essential ingredient in the design of information, communication,
and transportation networks, as well as in energy-related disciplines
such as power grid maintenance, control, and optimization \cite{Pinar2010}.
Graph theory and linear algebra provide abstractions and quantitative 
tools that can be employed in the analysis and design of large and
complex network models.

Real-world networks are characterized by structural properties 
that make them very different from both regular graphs on one hand, and  
completely random graphs on the other. Real networks frequently exhibit
a highly skewed degree distribution (often following a power law), small
diameter, high clustering coefficient (the two last properties together are
often referred to as the {\em small world} property), the presence of
motifs, communities, and other signatures of complexity.   

Some of the basic questions in network analysis concern node and edge centrality, 
community detection, communicability, and diffusion 
\cite{Brandes,EstradaBook,NewmanBook}. Related to these are the
important notions of network robustness (or its opposite, vulnerability) and 
connectivity \cite{Cohen}. These latter properties refer to the degree of resiliency
displayed by the network in the face of random accidental failures or 
deliberate, targeted attacks, which can be modeled in terms of edge or node
removal. Generally speaking, it is desirable to design networks that are at
the same time highly sparse (in order to reduce costs) and highly connected,
meaning that disconnecting or disrupting the network would require the removal of a large
number of edges. Such networks should not contain bottlenecks, and they
should allow for the rapid exchange of communication between nodes.
Expander graphs \cite{E06b,HLW06} are an important class of graphs 
with such properties. 

%Moreover, there are situations (such as counterterrorism 
%operations or the containment of infectious diseases) where one may actually
%want to disrupt as much as possible communication between nodes while modifying
%the network as little as possible. Clearly, this problem is closely connected
%to the previous one. 

In this paper we describe some techniques that can be brought to bear on the
problems described above and related questions.
Our approach is based on the notion of {\em total communicability} of a
network, introduced in \cite{Benzi2013} on the basis of earlier work
by Estrada and Hatano \cite{EH08,EHB12}. Total communicability, defined
as the (normalized) sum of the entries in the exponential of the adjacency
matrix of the network, provides a global measure of how well the nodes
in a graph can exchange information. 
Communicability is based on the number and length
of graph walks connecting pairs of nodes in the network. Pairs of nodes $(i,j)$
with high communicability correspond to large entries $[e^{A}]_{ij}$ in the
matrix exponential of $A$, the adjacency matrix of the network.

Total network communicability can also be used to measure the 
connectivity of the network as a whole. For instance, given two alternative
network designs (with a similar ``budget" in terms of number
of candidate edges), one can compare the two designs by computing the
respective total communicabilities and pick the network with the highest one,
assuming that a well-connected network with high node communicability is
the desired goal. 
It is important to stress that
the total communicability of a network can be efficiently computed or estimated
even for large networks using Lanczos or Arnoldi based algorithms without
having to compute any individual entry of $e^A$ (only the ability to perform
matrix-vector products with $A$ is required).

%In this paper we consider the following problems:
%{\bf Maybe summarize these a bit? No need to state them formally here, since we do it later in \S4}
%\begin{itemize}
%\item Given an existing connected (undirected) network $G=(V,E)$, choose an existing
%edge $(i,j)\in E$ such that removing this edge would minimize the decrease
%in the total communicability of the network (and preserve the network connectedness);
%this is repeated until a prescribed reduction in cost (measured in terms of
%deleted edges) has been achieved.
% We refer to this as the {\em downdating problem}.
%\item Given an existing sparse (undirected) network $G=(V,E)$, choose a pair of nodes
%$i,j\in V$ (with $i\ne j$ and $(i,j)\notin E$) such that adding the edge
%$(i,j)$ to $E$ would maximize the increase in the total communicability of the
%network; this is then repeated until a prescribed budget of new edges
%has been exhausted. We refer to this as the {\em updating problem}.
%\item Given an existing sparse connected (undirected) network $G=(V,E)$, perform a
%sequence of downdate-then-update steps,
%with the goal of maximizing the total
%communicability of the modified network; we call this the {\em rewiring problem}.
%\end{itemize} 

In this paper we consider three different problems. 
Let $G=(V,E)$ be a connected, undirected and sparse graph. 
The {\it downdating problem} consists of selecting an edge $(i,j)$ to 
be removed from the network so as to minimize the decrease in its 
total communicability while preserving its connectedness.

The goal when tackling the {\it updating problem}, on the other hand, is to select a pair of 
nodes $i\neq j$ such that $(i,j)\not\in E$ in such a way that the increase in 
the total communicability of the network is maximized. 

Finally, the {\it rewiring problem} has the same goal as the updating problem, but it requires 
the selection of two modifications which constitute the downdate-then-update step to be performed.

The importance of the first two problems for network analysis and
design is obvious.
We note that an efficient solution to the second problem would also 
suggest how to proceed if the goal was to identify existing edges whose
removal would {\em maximize} the decrease in communicability, which could
be useful, e.g., in planning anti-terrorism operations or public health policies 
(see, e.g., \cite{Tong, VM2011}).
The third problem is motivated by the observation that
for transportation networks (e.g., flight routes) it is sometimes 
desirable to redirect edges in order to improve the performance 
(i.e., increase the number of travellers) without increasing too much 
the costs. 
Hence, in such cases, one wants to eliminate a route used only by a few 
travellers and to add a route that may be used by a lot of people.

The above problems may arise not only in the design of infrastructural networks
(such as telecommunication or transportation networks), but also in other
contexts. For instance, in social networks the addition of a friendship/collaborative
tie may change dramatically the structure of the network, leading
to a more cohesive
group, and hence preventing the splitting of the community into smaller
subgroups.
 
%%Another possibility is that this addition may strengthen the existing 
%%smaller groups within the network, making them more resilient under 
%%further splitting once the network is broken into smaller subgraphs 
%%according to the existing structure.
%%Both of these behaviors (``shorten'' the network or strengthen the existing 
%%subgraphs) appear naturally in social networks.
%%Clearly, this second strategy is not advisable when designing an 
%%infrastructural or transport network, since 
%%it would lead to a network that can be easily split into subgraphs, i.e., 
%%a network that is not resilient under targeted attacks based, for example, 
%%on the betweenness centrality (see \cite{Girvan2002}). 
%%
%%This approach, namely the rewire, may be sometimes preferred to 
%%the update of an edge. 
%%It is composed of a downdating step that must not disconnect the 
%%network, nor isolate a node from the rest 
%%of the graph, and of an updating step. 
%%The first step should not penalize too much the ease of communicating, 
%%whilst the second step should increase as much as possible the 
%%value for the total 
%%communicability.
%%These two steps are aimed to the same goals described for the downdate 
%%and the update themselves and hence it seems reasonable to combine the strategies 
%%for these two to get a good rewiring technique.

%In this paper we develop updating, downdating, 
%and rewiring techniques which manipulate 
%the edges of the network in such a way that the total communicability 
%responds as desired to these modifications.

The work is organized as follows. 
Section \ref{sec:background} contains some basic facts from linear 
algebra and graph theory, and introduces the modifications of the adjacency 
matrix we will perform. 
 In this section we also provide further justification
for the use of the total network communicability as the objective
function.
In section \ref{sec:bounds}  we describe bounds for the 
total communicability 
via the Gauss--Radau quadrature rule and we show how these bounds 
change when a rank-two modification of the adjacency matrix is performed.
Section \ref{sec:modification} is devoted to the introduction of the 
methods to controllably modify the graph in order to adjust the value of its 
total communicability.
Numerical studies to assess the effectiveness and performance of the 
techniques introduced are provided in section \ref{sec:test_tc} for 
both synthetic and 
real-world networks.
In section \ref{sec:nc+gap} we discuss the evolution of 
a popular measure of network connectivity, known as
the {\em free energy} (or {\em natural connectivity}), 
when the same modifications are performed. 
This section provides further evidence that motivates the use of the total 
communicability as a measure of connectivity. 
Finally, in section \ref{sec:conclusions} we draw conclusions and we describe 
future directions.

%%%%%%%%%%%%%%%%%%%%%%%%%%%%%%%%%%%%%%%%%%%%%%%%%%%%%%%%%%%%%%%%%%%%%%%%%%
%%%%%%%%%%%%%%%%% 2 BACKGROUND AND DEFINITIONS %%%%%%%%%%%%%%%%%%%%%%%%%%%
%%%%%%%%%%%%%%%%%%%%%%%%%%%%%%%%%%%%%%%%%%%%%%%%%%%%%%%%%%%%%%%%%%%%%%%%%%
\section{Background and definitions}
\label{sec:background}
In this section we provide some basic
definitions, notations, and properties associated with graphs.

A {\itshape graph} or {\itshape network}
$G=(V,E)$ is defined by a set of $n$ nodes (vertices) 
$V$ and a set of $m$ edges $E=\{(i,j)|i,j\in V\}$ between the nodes. 
An edge is said to be {\itshape incident} to a vertex $i$ if there exists 
a node $j\neq i$ such that either $(i,j)\in E$ or $(j,i)\in E$.
The {\itshape degree} of a vertex, denoted by $d_i$, is the number of 
edges incident to $i$ in $G$. 
The graph is said to be {\itshape undirected} if the edges are formed by 
unordered pairs of vertices. 
A {\itshape walk} of length $k$ in $G$ is a set of nodes $i_1, i_2,\ldots,i_k, 
i_{k+1}$ such that for all $1\leq l\leq k$, $(i_l,i_{l+1})\in E$.
A {\itshape closed walk} is a walk for which $i_1=i_{k+1}$. 
A {\itshape path} is a walk with no repeated nodes. 
A graph is {\itshape connected} if there is a path connecting every pair of nodes.
A graph with unweighted edges, no self-loops (edges from a node to itself), 
and no multiple edges is said to be {\itshape simple}. 
Throughout this work, we will consider undirected, simple, and connected networks.

Every graph can be represented as a matrix 
$A=\left(a_{ij}\right)\in\mathbb{R}^{n\times n}$, called the 
{\itshape adjacency matrix} of the graph. 
The entries of the adjacency matrix of an unweighted graph $G=(V,E)$ are 
\begin{equation*}
a_{ij}=\left\{
\begin{array}{ll}
1 & \mbox{if } (i,j)\in E\\
0 & \mbox{otherwise}
\end{array}
\right.\qquad \forall i,j\in V.
\end{equation*}
If the network is simple, the diagonal elements of the adjacency matrix 
are all equal to zero.
In the special case of an undirected network, the associated adjacency matrix 
is symmetric, and thus its eigenvalues are real.

We label the eigenvalues in non-increasing order: 
$\lambda_1\geq\lambda_2\geq\cdots \geq \lambda_n$.
Since $A$ is a real-valued, symmetric matrix, we can decompose $A$ 
into $A=Q\Lambda Q^T$ where $\Lambda$ is a diagonal matrix containing the 
eigenvalues of $A$ and $Q=[\mathbf{q}_1,\ldots,\mathbf{q}_n]$ is orthogonal, 
where $\mathbf{q}_i$ is an eigenvector associated with $\lambda_i$.
Moreover, if $G$ is connected, $A$ is irreducible and from the 
Perron--Frobenius Theorem \cite[Chapter 8]{Meyer00} we deduce that $\lambda_1>\lambda_2$ 
and that the leading eigenvector $\mathbf{q}_1$, 
sometimes referred to as the {\itshape Perron vector}, 
can be chosen such that its components $q_1(i)$ are positive 
for all $i\in V$.

We can now introduce the basic operations which will be performed 
on the adjacency matrix $A$ associated with the network $G=(V,E)$.
We define the {\itshape downdating} of the edge $(i,j)\in E$ as the 
removal of this edge from the network.
The resulting graph $\widehat{G}=(V,\widehat{E})$, which may be disconnected, 
has adjacency matrix 
\begin{equation*}
\widehat{A}=A-UW^T, \qquad U=[\mathbf{e}_i,\mathbf{e}_j],
\quad W=[\mathbf{e}_j,\mathbf{e}_i],
\end{equation*}
where here and in the rest of this work  the vectors 
$\mathbf{e}_i$, $\mathbf{e}_j$ represent
the $i$th and $j$th vectors of the standard basis of $\mathbb{R}^n$, 
respectively.

Similarly, let $(i,j)\in\overline{E}$ be an element in the complement of $E$. 
We will call this element a {\itshape virtual edge} for the graph $G$. 
We can construct a new graph $\tilde{G}=(V,\tilde{E})$ obtained from $G$ 
by adding the virtual edge $(i,j)$ to the graph.
This procedure will be referred to as the {\itshape updating} of the 
virtual edge $(i,j)$.
The adjacency matrix of the resulting graph is
\begin{equation*}
\tilde{A}=A+UW^T, \qquad U=[\mathbf{e}_i,\mathbf{e}_j],
\quad W=[\mathbf{e}_j,\mathbf{e}_i].
\end{equation*}
Hence, these two operations can both be described as 
rank-two modifications of the adjacency matrix of the original graph.

The operation of downdating an edge and successively updating a virtual 
edge will be referred to as {\itshape rewiring}. 
\begin{rem}
{\rm These operations are all performed in a symmetric fashion, since in this paper we 
consider exclusively undirected networks.}
\end{rem}

\subsection{Centrality and total communicability}
\label{subsec:centrality}
%%Networks model complex system from a wide range of disciplines. 
One of the main goals when analyzing a network is to identify the
most influential nodes in the network. 
Over the years, various measures of the importance, or centrality, 
of nodes have been developed \cite{Brandes,EstradaBook,NewmanBook}. 
In particular the {\itshape (exponential) subgraph centrality} 
of a node $i$ (see \cite{Estrada2005}) is defined as the $i$th diagonal 
element of the matrix exponential \cite{Higham2008}:
$$e^A=I+A+\frac{A^2}{2!}+\ldots=\sum_{k=0}^\infty\frac{A^k}{k!},$$
where $I$ is the $n\times n$ identity matrix.
As it is well known in graph theory, given an adjacency matrix $A$ 
of an unweighted network and $k\in\mathbb{N}$, 
the element $\left(A^k\right)_{ij}$ counts the total number of walks 
of length $k$ starting from node $i$ and ending at node $j$.
Therefore, the subgraph centrality of node $i$ counts the total number 
of closed walks centered at node $i$, weighting walks of length $k$ by a factor 
$\frac{1}{k!}$, hence giving more importance to shorter walks. 
The subgraph centrality then accounts for the returnability of the 
information to the node which was the source of this same information. 
Likewise, the off-diagonal entries of the adjacency matrix 
$\left(e^A\right)_{ij}$ ({\itshape subgraph communicability} of nodes $i$ and $j$) 
account for the ability of nodes $i$ and $j$ to exchange information 
\cite{EH08,EHB12}.

Starting from these observations and with the aim of reducing the cost 
of the computation of the rankings, in \cite{Benzi2013} it was suggested to use
as a centrality measure 
the \emph{total communicability of a node} $i$, defined as
the $i$th entry of the vector
$e^A\mathbf{1}$, where $\mathbf{1}$ denotes the vector of all ones:
\begin{equation}\label{tnc}
TC(i):= [e^A\mathbf{1}]_i = \sum_{j=1}^n \left[e^A\right]_{ij}.
\end{equation}
This measure of centrality is given by a weighted sum of walks from
every node in the network (including node $i$ itself), and thus quantifies
both the ability of a node to
spread information across the network and the returnability of the information to 
the node itself.

The value resulting from summing these quantities over all the nodes %, 
%i.e.,  $\mathbf{1}^Te^A\mathbf{1}$, 
can be interpreted as a global measure of how effectively the 
communication takes place across the whole network.
This index is called {\itshape total (network) communicability} 
\cite{Benzi2013} and can be written as
\begin{equation}\label{tc_spec}
TC(A):=\mathbf{1}^Te^A\mathbf{1}=\sum_{i=1}^n\sum_{j=1}^n(e^A)_{ij} = 
\sum_{k=1}^n e^{\lambda_k}({\bf q}_k^T{\bf 1})^2.
%\sum_{i=1}^n\sum_{j=1}^n\sum_{k=1}^n e^{\lambda_k}q_k(i)q_k(j). 
\end{equation} 
This value can be efficiently computed, e.g., by means of a 
Krylov method as implemented in S.~G\"uttel's Matlab toolbox \texttt{funm\_kryl} 
see \cite{Krylov1,Guettel} or by Lanczos-based techniques 
as discussed below. In the toolbox \cite{Guettel}
an efficient algorithm for evaluating $f(A)\mathbf{v}$ is implemented; 
with this method the vector $e^A\mathbf{1}$ can be constructed in 
roughly $O(n)$ operations (note that the prefactor can vary %considerably 
for different types of networks) and the total communicability is easily derived.

As it is clear from its definition, the value of $TC(A)$ may be very large. 
Several normalizations have been proposed; the simplest is the normalization 
by the number of nodes $n$ (see \cite{Benzi2013}), which we 
will use throughout the paper.
It is easy to prove that the normalized 
total communicability satisfies
\begin{equation}\label{coarse_bounds}
\frac{1}{n}\sum_{i=1}^n\left(e^A\right)_{ii}\leq \frac{TC(A)}{n}\leq e^{\lambda_1},
\end{equation}
where the lower bound is attained by the graph with $n$ nodes and no 
edges and the upper bound is attained by the complete graph with $n$ nodes. 

\begin{rem}\label{rem:lambda1}
{\rm The last equality in equation \eqref{tc_spec} shows that 
the main contribution to the value of $TC(A)$ is likely to come from 
the term $e^{\lambda_1}\|{\bf q}_1\|_1^2$.} 
\end{rem}

\subsection{Rationale for targeting the total communicability} \label{why_TM}
As already mentioned, the total communicability provides a good measure
of how efficiently information (in the broad sense 
of the term) is diffused across the network. Typically, very
high values of $TC(A)$ are observed
for highly optimized infrastructure networks 
(such as airline routes or computer networks) and for highly cohesive social and
information networks (like certain type of collaboration networks). 
Conversely, the total network communicability is relatively low for spatially
extended, grid-like networks (such as many road networks) or for 
networks that consist of two or more communities with poor communication
between them (such as the Zachary network).\footnote{ Numerical
values of the normalized total network communicability for a broad
collection of networks are reported in
the experimental sections of this paper, in the Supplementary Material,
and in \cite{Benzi2013}.}
As a further example,
reduced values of the communicability between different brain regions have
been detected in stroke patients compared to healthy individuals 
\cite{CH09}.
We refer to \cite{EHB12} for an extensive survey on communicability,
including applications for which it has been found to be useful.

Another reason in support of the use of the total communicability as
an objective function is that it is closely related to the {\em natural
connectivity} (or {\em free energy}) of the network, while being
dramatically easier to compute; see section \ref{sec:nc+gap} below.  
Sparse networks with high
values of $TC(A)$ are very well connected and thus less likely to 
be disrupted by either random failures or targeted attacks leading
to the loss of edges. This justifies trying to design sparse networks with
high values of the total communicability.

An important observation is that the total network communicability $TC(A)$
can be interpreted in at least two different ways. Since it is given by the
sum of all the pairwise communicabilities $C(i,j)=[e^A]_{ij}$, it is a
global measure of the ability of the network to diffuse information.
However, recalling the definition (\ref{tnc}) of total node communicability,
the normalized total communicability can also be seen as ``the average
total communicability" of the nodes in the network: 
$$\frac{TC(A)}{n} = \frac{1}{n}\sum_{i=1}^n TC(i).$$
Since the total node
communicability is a centrality measure \cite{Benzi2013}, our goal  
can then be rephrased as the problem of constructing sparse networks
having high average node centrality, where the node centrality is given
by total node communicability. Since this is merely one of a large number
of centrality measures proposed in the literature, it is a legitimate 
question to ask why the total node communicability should be used instead of
a different centrality index.  In other words, given any node
centrality function $f:V \longrightarrow \mathbb R_+$, we could consider 
instead the problem of, say, adding a prescribed number of edges
so as to maximize the increase in the global average centrality
$$\bar f = \frac{1}{n}\sum_{i=1}^n f(i).$$

As it turns out, most other centrality indices are either computationally
too expensive to work with (at least for large networks), or lead to objective
functions which do not make much sense. The following is a brief discussion
of some of the most popular centrality indices used in the field of
network science.

\vspace{0.1in}

\begin{enumerate}
\item {\bf Degree:} Consider first the simplest centrality index, the degree. 
Obviously, adding $K$ edges according to {\em any} criteria will
produce exacty the same variation in the average degree of a network.
Hence, one may as well add edges at random. Doing so, however,
cannot be expected to be greatly beneficial if the goal is to 
improve the robustness or efficiency of the network.
\item {\bf Eigenvector centrality:} Let ${\bf q}_1$ be the principal
eigenvector of $A$, normalized so that $\|{\bf q}_1\|_2 = 1$. The eigenvector
centrality of node $i\in V$ is the $i$th component of ${\bf q}_1$, denoted
by $q_1(i)$. It is straightforward to see that the problem of maximizing
the average eigenvector centrality
$$\frac{q_1(1) + q_1(2) + \cdots + q_1(n)}{n}$$
subject to the constraint $\|{\bf q}_1\|_2 = 1$ has as its only solution
$$q_1(1) = q_1(2) = \cdots = q_1(n) = \frac{1}{\sqrt n}.$$
This implies that $A$ has constant row sums or, in other words, that the
graph is regular --- every node in $G$ has the same degree. Hence, any
heuristic aimed at maximizing the average eigenvector centrality will
result in graphs that are close to being regular. However, regular 
graph topologies are not, {\em per se}, endowed with any especially good  
properties when it comes to diffusing information or being robust: think
of a cycle graph, for example. Regular graphs {\em can} be very well connected
and robust (this is the case of
expander graphs), but there is no reason to think that
simply making the degree distribution of a given network more regular will
improve its expansion properties.
\item {\bf Subgraph centrality}: the average subgraph centrality of a
network is known in the literature as the normalized {\em Estrada index}:
$$\frac{1}{n}EE(A) = \frac{1}{n}{\text Tr} (e^A) = 
\frac{1}{n}\sum_{i=1}^n [e^A]_{ii} = \frac{1}{n}\sum_{i=1}^n 
e^{\lambda_i}.$$
It can also be interpreted as the average self-communicability of the
nodes. As we mentioned, this is a lower bound for the average total
communicability.  Evaluation of this quantity requires knowledge
of all $n$ diagonal entries of $e^A$, or of all the eigenvalues of $A$ 
and is therefore much more expensive to compute. The heuristics we
derive in this paper have a similar effect on $TC(A)$ and on the
Estrada index, as we demonstrate in section \ref{sec:nc+gap}. So, 
using subgraph centrality instead of total communicability centrality
would lead to exactly the same heuristics and results, with the
disadvantage that evaluating the objective function, if necessary, would
be much more expensive.
\item {\bf Katz centrality}: the Katz centrality of node $i\in V$ is
defined as the $i$th row sum of the matrix resolvent
 $(I - \alpha A)^{-1}$, where
the parameter $\alpha$ is chosen in the interval $(0,\frac{1}{\lambda_1})$,
so that the power series expansion
$$(I - \alpha A)^{-1} = I + \alpha A + \alpha^2 A^2 + \cdots$$
is convergent \cite{Katz}.
Since this centrality measure can be interpreted in terms of walks,
using it instead of the total communicability
would lead to the same heuristics and very similar
results, especially when
$\alpha$ is sufficiently close to $\frac{1}{\lambda_1}$ or if the
spectral gap $\lambda_1 - \lambda_2$ is large; see \cite{Benzi2015}.
Using Katz centrality, however, requires the careful selection of the
parameter $\alpha$, which leads to some complications. For example,
after each update one needs to recompute the dominant eigenvalue
of the adjacency matrix in order to check whether the value of
$\alpha$ used is still within the range of permissible values or if
it has to be reduced, making
this approach computationally very expensive. This
problem does not arise if the matrix exponential is used instead
of the resolvent.  
\item {\bf Other centrality measures}: So far we have only discussed
centrality measures that can be expressed in terms of the adjacency matrix $A$.
These centrality measures are all connected to the notion of walk in a graph,
and they can often be understood in terms of spectral graph theory.
Other popular centrality measures, such as betweenness centrality and
closeness centrality (see, e.g., \cite{NewmanBook}) do not have a simple
formulation in terms of matrix properties. They are based on the assumption
that all communication in a graph tends to take place along shortest paths,
which is not always the case (this was a major motivation for the
introduction of walk-based measures, which postulate that communication
between nodes can take place along walks of any length, with a preference
towards shorter ones). A further disadvantage is that they are quite
expensive to compute, although randomized approximations can bring the
cost down to acceptable levels \cite{Brandes}. For these reasons we do not
consider them in this paper, where the focus is on linear algebraic
techniques. It remains an open question whether heuristics  for
manipulating graph edges so as to tune some gloabl average of these
centrality measures can lead to networks with desirable connectivity
and robustness properties.
\end{enumerate}

\vspace{0.1in}

Finally, in view of the bounds (\ref{coarse_bounds}), the evolution of
the total communicability under network modifications is closely tied
to the evolution of 
the dominant eigenvalue $\lambda_1$. This quantity plays a
crucial role in network analysis, for example in the definition
of the {\em epidemic threshold}; see, for instance, \cite[p.~664]{NewmanBook} and
\cite{VM2011}. In particular, a decrease in the total network 
communicability can be expected to lead to an increase in the
epidemic threshold. 
Thus, edge modification techniques developed for tuning $TC(A)$
can potentially be used to
alter epidemics dynamics.

%%%%%%%%%%%%%%%%%%%%%%%%%%%%%%%%%%%%%%%%%%%%%%%%%%%%%%%%%%%%%%%%%%%%%%%%%%
%%%%%%%%%%%%%%%%%% 3 BOUNDS VIA QUADRATURE RULES %%%%%%%%%%%%%%%%%%%%%%%%%
%%%%%%%%%%%%%%%%%%%%%%%%%%%%%%%%%%%%%%%%%%%%%%%%%%%%%%%%%%%%%%%%%%%%%%%%%%
\section{Bounds via quadrature rules}
\label{sec:bounds}
In the previous section we saw the simple bounds (\ref{coarse_bounds}) on
the normalized total network communicability. 
More refined bounds for this index can be obtained by means of quadrature 
rules as described in \cite{Benzi2010,Benzi1999,Golub2010,Fenu2013}.
The following theorem contains our result on the bounds for the normalized total communicability.
\begin{theorem}\label{thm:bounds}
Let $A$ be the adjacency matrix of an unweighted and undirected network. Then
\begin{equation*}
\Phi\left(\beta,\omega_1+\frac{\gamma_1^2}{\omega_1-\beta}\right)\leq 
\frac{TC(A)}{n}\leq\Phi\left(\alpha,\omega_1+\frac{\gamma_1^2}{\omega_1-\alpha}\right)
\end{equation*}
where $[\alpha,\beta]$ is an interval containing the spectrum of $-A$ 
(i.e., $\alpha \le -\lambda_1$ and $\beta\ge -\lambda_n$), 
$\omega_1=-\mu=-\frac{1}{n}\sum_{i=1}^nd_i$ is the negative mean of the degrees, 
$\gamma_1=\sigma=\sqrt{\frac{1}{n}\sum_{k=1}^n(d_k-\mu)^2}$ is the standard deviation, and
\begin{equation}\label{eq:bound}
\Phi(x,y)=\frac{c \left(e^{-x}-e^{-y}\right)+xe^{-y}-ye^{-x}}{x-y}, \qquad c=\omega_1. 
\end{equation}
\end{theorem}

A proof of this result can be found in the Supplementary Materials 
accompanying the paper.

%Following the same procedure, 
Analogous bounds can be found for the 
adjacency matrix of the graph after performing 
a downdate or an update. 
These results are summarized in the following Corollaries.

\begin{corollary}\label{Dwdt} [Downdating]
Let $\widehat{A}=A-UW^T$, where $U=[\mathbf{e}_i,\mathbf{e}_j]$ and
$W=[\mathbf{e}_j,\mathbf{e}_i]$ be the adjacency matrix of an unweighted and 
undirected network obtained after the downdate of the edge $(i,j)$ from the matrix $A$.
Let $\omega_1=-\mu=-\frac{1}{n}\sum_{i=1}^nd_i$ and $\gamma_1=\sigma=
\sqrt{\frac{1}{n}\sum_{k=1}^n(d_k-\mu)^2}$, where $d_i$ is the degree of node $i$ in the 
original graph. 
Then
\begin{equation*}
\Phi\left(\beta_{-},\omega_{-}+\frac{\gamma_{-}^2}{\omega_{-}-\beta_{-}}\right)
\leq\frac{TC(\widehat{A})}{n}\leq
\Phi\left(\alpha_-,\omega_-+\frac{\gamma_-^2}{\omega_--\alpha_-}\right)
\end{equation*}
where
\begin{equation*}
\left\{
\begin{array}{l}
\omega_{-}=\omega_1+\frac{2}{n};\\[6pt]
%\gamma_{-}=\left[\gamma_1^2-\frac{2}{n}\left(d_i+d_j-1+2\omega_1+
%\frac{2}{n}\right)\right]^{\frac{1}{2}}
\gamma_{-}=\sqrt{\gamma_1^2-\frac{2}{n}\left(d_i+d_j-1+2\omega_1+
\frac{2}{n}\right)}
\end{array}
\right. ,
\end{equation*}
$\alpha_-$ and $\beta_-$ are approximation of the smallest and largest 
eigenvalues of $-\widehat{A}$ respectively, 
and $\Phi$ is defined as in equation \eqref{eq:bound} with $c=\omega_-$.
\end{corollary}

Note that if bounds $\alpha$ and $\beta$ for the extremal eigenvalues of 
the original matrix are known, we can then use $\alpha_-=\alpha$ and $\beta_-=\beta+1$.
Indeed, if we order the eigenvalues of $\widehat{A}$ in non--increasing order 
$\widehat{\lambda}_1>\widehat{\lambda}_2\geq \cdots \geq\widehat{\lambda}_n$ we
obtain, as a consequence of Weyl's Theorem
(see \cite[Section 4.3]{Horn}), that
\begin{equation*}
\alpha-1 \leq -\lambda_1-1 < -\widehat{\lambda}_1 < 
-\widehat{\lambda}_2 \leq \cdots \leq -\widehat{\lambda}_n < -\lambda_n+1 \leq \beta+1.
\end{equation*}

Furthermore, the Perron--Frobenius Theorem ensures that, when performing a 
downdate, the largest eigenvalue of the adjacency matrix cannot increase; 
hence, we deduce the more stringent bounds $\alpha\leq-\widehat{\lambda}_1\leq
-\widehat{\lambda}_2\leq\cdots\leq -\widehat{\lambda}_n\leq \beta +1.$

Similarly, we can derive bounds for the normalized total communicability 
of the matrix $\tilde{A}$ 
obtained from the matrix $A$ after performing the update of the virtual edge $(i,j)$.

\begin{corollary}\label{Updt} [Updating]
Let $\tilde{A}=A+UW^T$, where $U=[\mathbf{e}_i,\mathbf{e}_j]$ and
$W=[\mathbf{e}_j,\mathbf{e}_i]$ be the adjacency matrix of an unweighted and 
undirected network obtained after the update of the virtual edge 
$(i,j)$ in the matrix $A$.
Let $\omega_1=-\mu=-\frac{1}{n}\sum_{i=1}^nd_i$ and $\gamma_1=\sigma=
\sqrt{\frac{1}{n}\sum_{k=1}^n(d_k-\mu)^2}$, where $d_i$ is 
the degree of node $i$ in the 
original graph.
Then

\begin{equation*}
\Phi\left(\beta_{+},\omega_{+}+\frac{\gamma_{+}^2}{\omega_{+}-\beta_{+}}\right)\leq 
\frac{TC(\tilde{A})}{n}
\leq\Phi\left(\alpha_+,\omega_++\frac{\gamma_+^2}{\omega_+-\alpha_+}\right)
\end{equation*}
where
\begin{equation*}
\left\{
\begin{array}{l}
\omega_{+}=\omega_1-\frac{2}{n};\\[6pt] 
%\gamma_{+}=\left[\gamma_1^2+\frac{2}{n}\left(d_i+d_j+1+2\omega_1-
%\frac{2}{n}\right)\right]^{\frac{1}{2}}
\gamma_{+}=\sqrt{\gamma_1^2+\frac{2}{n}\left(d_i+d_j+1+2\omega_1-\frac{2}{n}\right)}
\end{array}
\right. ,
\end{equation*}
$\alpha_+$ and $\beta_+$ are bounds for the smallest and largest eigenvalues of 
$-\tilde{A}$ respectively, and $\Phi$ is defined as in equation 
\eqref{eq:bound} with $c=\omega_+$.
\end{corollary}

Notice that again, if bounds $\alpha$ and $\beta$ for the extremal 
eigenvalues of $-A$ are known, we can then take $\alpha_+=\alpha-1$ and $\beta_+=\beta$.
In fact, the spectrum of the rank-two symmetric perturbations $UW^T$ and $-UW^T$ is 
$\{\pm 1,0\}$ and hence we can 
use Weyl's Theorem as before and then improve the upper bound
using the Perron--Frobenius Theorem.

In the next section we will see how the new bounds can be used to 
guide the updating and downdating process.

%%%%%%%%%%%%%%%%%%%%%%%%%%%%%%%%%%%%%%%%%%%%%%%%%%%%%%%%%%%%%%%%%%%%%%%%%%
%%%%%%%%%%%%%%%% 4 MODIFICATIONS OF THE ADJACENCY MATRIX %%%%%%%%%%%%%%%%%
%%%%%%%%%%%%%%%%%%%%%%%%%%%%%%%%%%%%%%%%%%%%%%%%%%%%%%%%%%%%%%%%%%%%%%%%%%

\section{Modifications of the adjacency matrix}
\label{sec:modification}
In this section we develop techniques that allow us to tackle 
the following problems.

\begin{itemize}
\item[(P1)] Downdate: select $K$ edges that can be downdated from 
the network without disconnecting it and that cause the smallest
drop in the total 
communicability of the graph;
\item[(P2)] Update: select $K$ edges to be added to the network 
(without creating self--loops or multiple edges) so as to increase as much as 
possible the total communicability of the graph;
\item[(P3)] Rewire: select $K$ edges to be rewired in the network 
so as to increase as much as possible the value of $TC(A)$. The
rewiring process must not disconnect the network or 
create self--loops or multiple edges in the graph.
\end{itemize}

As we will show below, (P3) can be solved using combinations of methods 
developed to solve (P1) and (P2).
Hence, we first focus on the downdate and  the update separately. 
Note that to decrease as little as possible the total communicability 
when removing an edge it would suffice to select $(i^*,j^*)\in E$ so as
to minimize the quantities
$$\mathbf{1}^TA^k\mathbf{1} -
\mathbf{1}^T(A-UW^T)^k\mathbf{1}\qquad \forall k=1,2,\ldots ,$$
since $TC(A)=\sum_{k=0}^\infty\frac{\mathbf{1}^TA^k\mathbf{1}}{k!}$.
Similarly, to increase as much as possible $TC(A)$ by addition of a virtual edge, 
it would suffice to select $(i^*,j^*)\in\overline{E}$ 
that maximizes the differences 
$$\mathbf{1}^T(A+UW^T)^k\mathbf{1}-\mathbf{1}^TA^k\mathbf{1}\qquad \forall k=1,2,\ldots $$
However, it is easy to show that in general one can not find a choice for
 $(i^*,j^*)$ that works for all such $k$. 
Indeed, numerical experiments on small synthetic graphs   
(not shown here) show that in general the optimal edge selection 
for $k=2$ is different from the one for $k=3$.
Because of this, 
it is unlikely that one can find a simple
``closed form solution" to the problem, and
we need to develop approximation techniques.

The majority of the heuristics we will develop are based on new edge centrality measures. 
The idea underlying these is that it seems reasonable to assume that an edge is more likely used as communication channel if its adjacent 
nodes are given a lot of information to spread. 
We thus introduce three new centrality measures for edges based on this principle: edges connecting important nodes are themselves 
important. 

%The definitions are all based on well-known nodes centrality measures (see \cite{Estrada2005, NewmanBook, Benzi2013}) 
%and are defined for the elements of both $E$ and  $\overline{E}$.

\begin{definition}
For any $i,j\in V$ we define the {\rm edge subgraph centrality} of 
an existing/virtual edge $(i,j)$ as
\begin{equation}
^eSC(i,j)=\left(e^A\right)_{ii}\left(e^A\right)_{jj}.
\label{eq:edge_subgraph}
\end{equation}
\label{def:subgraph}
\end{definition}

This definition, based on the subgraph centrality of nodes, exploits the fact that the matrix exponential is symmetric positive definite 
and hence $(e^A)_{ii}(e^A)_{jj}>(e^A)_{ij}^2$. 
Therefore, the diagonal elements of $e^A$ somehow control its off-diagonal entries, 
hence they may contain 
enough information to infer the ``payload'' of the edges or of the virtual edges of interest. 
%As we will see in section \ref{sec:test_tc}, this measure is by far the most computationally expensive among those we tested, 
%since it requires the computation of the diagonal entries of $e^A$.  
%However, in principle the cost could be greatly reduced using parallel processing, since each diagonal element can be computed independently of the 
%others. 
%{\bf Need a bridge here.}

\begin{definition}
For any $i,j\in V$ we define the {\rm edge total communicability centrality}
of an existing/virtual
edge $(i,j)$ as
\begin{equation}
^eTC(i,j) = [e^A{\bf 1}]_i [e^A{\bf 1}]_j.
\end{equation}
\label{def:tc_edge}
\end{definition}

It is important to observe that when
the spectral gap $\lambda_1-\lambda_2$ is ``large enough'', 
then the subgraph centrality $\left(e^A\right)_{ii}$ 
and the total communicability centrality $[e^A{\bf 1}]_i$ are
essentially 
determined by 
$e^{\lambda_1}q_1(i)^2$ and $e^{\lambda_1}q_1(i)\|{\bf q}_1\|_1$, respectively (see, e.g., \cite{Benzi2013,Benzi2015,E06b}); it 
follows that in this case 
the two centrality measures introduced and a centrality measure based on the eigenvector centrality for nodes can be expected to provide similar rankings. 
This is especially true when attention is restricted to the top edges (or nodes).
This observation motivates the introduction of the following edge centrality measure.

\begin{definition}
For any $i,j\in V$ we define the {\rm edge eigenvector centrality} of an existing/virtual 
edge $(i,j)$ as
\begin{equation}
^eEC(i,j)=q_1(i)q_1(j).
\label{eq:edge_eigenvector}
\end{equation}
\label{def:eigenvector}
\end{definition}
As a further justification for this definition, note that 

$$\lambda_1-2 \left( ^eEC(i,j)\right)\leq\widehat{\lambda}_1\leq\lambda_1,\qquad
\tilde{\lambda}_1\geq\lambda_1+2 \left( ^eEC(i,j)\right),$$

where $\widehat{\lambda}_1$ is the leading eigenvalue of the matrix $\widehat{A}$ and $\tilde{\lambda}_1$ 
is the leading eigenvalue of the matrix $\tilde{A}$, as defined in 
section \ref{sec:background}.
These inequalities show that the edge eigenvector centrality 
of an existing/virtual edge $(i,j)$ is strictly connected 
to the change in the value of the leading eigenvalue of the adjacency matrix, 
which influences
the evolution of the total communicability when we modify $A$ (see Remark \ref{rem:lambda1}). 

\begin{rem}
{\rm The edge eigenvector centrality has been used in \cite{Tong, VM2011} to 
devise edge removal techniques aimed to reduce significantly $\lambda_1$, so as 
to increase the {\em epidemic threshold} of networks.}
\end{rem}

Note that we defined these measures of centrality for 
both existing and virtual edges (as in \cite{Berry2013}). 
The reason for this as well as the justification for these definitions 
will become clear in the next subsections. 

%Given these definitions, we can infer that an edge having a low centrality is expected to carry only a small amount of information 
%and hence the network would not suffer too much from its removal. 
%On the other hand, a virtual edge which has a high centrality is expected to potentially be a good communication channel. 
%Hence, its addition is expected to highly enhance the overall ability of the network of spreading information.

We now discuss
 how to use these definitions to tackle the problems previously described. 
The computational aspects concerning the implementation of the heuristics we are about to introduce 
and the derivation of their computational costs are described in the 
Supplementary Materials. 

\subsection*{(P1) Downdate}
The downdate of any edge in the network will result in a reduction of 
its total communicability. 
%Since the total communicability is an index of the ease of sending 
%information across the network, our main goal when 
%performing a downdate is to reduce this index as little as possible. 
% the decrease in this index.
Note that since we are focusing on the case of connected networks, we 
will only perform downdates that keep the resulting graph connected.
In practice, it is desirable to further
restrict the choice of downdates to a subset of
all existing edges, on the basis of criteria to be discussed shortly.

An ``optimal" approach would select at each step of the downdating 
process a candidate edge corresponding to the minimum decrease of 
communicability.\footnote{Strictly speaking, this would correspond to
a greedy algorithm which is only locally optimal. In general, this
is unlikely to result in ``globally optimal" network communicability.
%%Because in this paper
%%we stipulate that only a single downdate, update, or rewire can be performed
%%at a given time step, we omit the qualifier ``locally".}
In this paper, the term ``optimal" will be understood in this limited
sense only.} 
Note that for large networks this method is too costly to be practical.
For this reason we aim to develop inexpensive techniques that will hopefully
give close--to--optimal results.
Nevertheless, for small networks we will use the
``optimal" approach (where we systematically try all feasible
edges and delete the one causing the least drop in total communicability) 
as a baseline method against which we compare the various
algorithms discussed below. This method will be henceforth
referred to as {\tt optimal}.

The next methods we introduce perform the downdate of the lowest ranked existing edge according to 
the edge centrality measures previously introduced whose removal does not disconnect the network. 
We will refer to these methods as {\tt subgraph}, {\tt nodeTC}, and {\tt eigenvector}, which are 
based on definitions \ref{def:subgraph}, \ref{def:tc_edge}, and \ref{def:eigenvector}, respectively. 
%The next method we introduce performs the downdate of the lowest 
%ranked existing edge according to the edge subgraph centrality whose removal does not 
%disconnect the network. We will refer to this method as {\tt subgraph}.
From the point of view of the communicability, %{\tt subgraph} downdates an 
these methods downdate an edge connecting two nodes which are peripheral (i.e., have low centrality)
and therefore are not expected to 
give a large contribution to the spread of information along the network.
%As further justification for this approach, we observe that
%\begin{equation}
%\left(e^A\right)_{ii}\left(e^A\right)_{jj}> \left(e^A\right)_{ij}^2\qquad \forall i,j\in V,
%\label{eq:disuguaglianza}
%\end{equation}
%since $e^A$ is positive definite.
Hence, the selected edge is connecting two nodes whose ability to
exchange information is already very low, and
we do not expect the total communicability to suffer too much from this edge removal.
This observation also suggests that such downdates 
can be repeatedly applied  without the need to recompute the ranking
of the edges after each downdate. 
As long as the number of downdates performed remains small compared to the total number of edges, 
we expect good results at a greatly reduced total cost. 
Note also that such downdates can be performed simultaneously rather than sequentially.
%We will refer to this variant as {\tt subgraph.no}.
We will refer to these variants as {\tt subgraph.no}, {\tt nodeTC.no}, and {\tt eigenvector.no}.

Finally, we consider a technique motivated by the bounds 
obtained via quadrature rules derived in section \ref{sec:bounds}.
From the expression for the function $\Phi$ in the special case of the downdate 
(cf.~Corollary \ref{Dwdt}),
we infer that a potentially good choice may be to remove the edge
having incident nodes $i,j$ for which the sum $d_i+d_j$ is minimal, if its 
removal does not disconnect the network.
Indeed, this choice reduces the upper bound only slightly and the total 
communicability may mirror this behavior.
Another way to justify this strategy is to observe that it is indeed
the optimal strategy if we approximate $e^A$ with its second-order
approximation $I + A + \frac{1}{2}A^2$ in the definition of total
communicability.
This technique will be henceforth referred to as {\tt degree}. 
We note that a related measure, namely, the average 
of the out-degrees $\frac{d_i+d_j}{2}$, 
was proposed in \cite{Berry2013} as a measure for the centrality of 
an edge $(i,j)$ in directed graphs.

\subsection*{(P2) Update}
Most real world networks are characterized by low average degree. 
As a consequence, the adjacency matrices of such networks are sparse ($m=O(n)$). 
For the purpose of selecting a virtual edge to be updated, this implies 
that we have approximately $\frac{1}{2}\left(n^2-cn\right)$ possible choices  
if we want to avoid the formation of multiple edges or self--loops (here $c$ is
a moderate constant).
Each one of these possible updates will result in an increase of the total 
communicability of the network, but not every one of these will result in a
significant increment. 

One natural updating technique is to connect two nodes having high 
%subgraph 
centralities, i.e., add the virtual 
edge having the highest ranking according to the corresponding edge centrality. 
Its incident nodes, being quite central, can be expected to have an 
important role in the spreading of information along the network; 
on the other hand, the communication between them may be relatively poor 
(think for example of the case where the two nodes sit in two distinct
communities). 
Hence, giving them a preferential communication channel, such as an 
edge between them, should result in a better spread of information along the 
whole network. 
%Again, we will use the label {\tt subgraph} to describe
%this updating strategy.
Again, we will use the labels {\tt subgraph}, {\tt nodeTC}, and {\tt eigenvector} to describe these 
updating strategies. 
As before, in order to reduce the computational cost, we also test the effectiveness 
of 
%this technique 
these techniques
without the recomputation of the 
ranking of the virtual edges after each update.
%This variant (referred to as {\tt subgraph.no})is 
These variants (referred to as {\tt subgraph.no}, {\tt nodeTC.no}, and {\tt eigenvector.no}) are 
expected to return good results as well, since 
the selected update should not radically change the ranking of the edges. 
Indeed, they make central nodes even more central, and the ranking of the 
edges remains consequently almost unchanged.
Note again that these updates can be performed simultaneously 
rather than sequentially.

%As for the case of downdating, less expensive methods result if
%eigenvector centrality or node total communicability
%are used instead of subgraph centrality.
%Hence, we also consider adding the virtual edge having the largest 
%edge eigenvector or edge total communicability
%centrality. We expect the corresponding results to
%be comparable to those obtained with edge subgraph centrality when
%the spectral gap is not too small. As before, the corresponding updating
%techniques (with and without recalculation of the node centralities)
%will be referred to as {\tt eigenvector}/{\tt nodeTC} 
%and {\tt eigenvector.no}/{\tt nodeTC.no},
%respectively.

As for the case of downdating, 
the bounds via quadrature rules derived in section \ref{sec:bounds} 
suggest an updating technique, i.e., adding the virtual edge $(i,j)$ for which $d_i+d_j$ is maximal. Indeed,
such a choice would maximize the lower bound on the total communicability, see
Corollary \ref{Updt}. Again, this choice can also be justified by noting that it
is optimal if $e^A$ is replaced by its quadratic Maclaurin approximant.
We will again use the label {\tt degree} to refer to this updating strategy.

All these techniques will be compared with the {\tt optimal} one, 
based on systematically trying all feasible virtual edges and selecting
at each step the one resulting in the largest increase of the
total communicability. Due to the very high cost of this brute force
approach, we will use it only on small networks.

\begin{table}
\footnotesize
\centering
\caption{Brief description of the techniques introduced in the paper.}
\begin{tabular}{lcc}
\hline
    Method                 & Downdate: $(i,j)\in E$ & Update: $(i,j)\not\in E$\\
\hline
    {\tt optimal}          & $\arg\min\{TC(A)-TC(\widehat{A})\}$ & $\arg\max\{TC(\tilde{A})-TC(A)\}$ \\
    {\tt subgraph(.no)}    & $\arg\min\{^eSC(i,j)\}$             & $\arg\max\{^eSC(i,j)\}$\\
    {\tt eigenvector(.no)} & $\arg\min\{^eEC(i,j)\}$             & $\arg\max\{^eEC(i,j)\}$\\
    {\tt nodeTC(.no)}      & $\arg\min\{^eTC(i,j)\}$             & $\arg\max\{^eTC(i,j)\}$\\
    {\tt degree}           & $\arg\min\{d_i+d_j\}$               & $\arg\max\{d_i+d_j\}$   \\
\hline
\end{tabular}
\label{tab:description}
\end{table}

The heuristics introduced to tackle (P1) and (P2) are summarized in Table \ref{tab:description}.

\subsection*{(P3) Rewire}
\label{subsec:rewire}
As we have already noted, there are situations in which
the rewire of an edge may be preferable 
to the addition of a new one.
There are various possible choices for the rewiring strategy to follow. 
The greatest part of those found in literature are variants of random rewiring 
(see for example \cite{Beygelzimer2005,Louzada2013}). 
In this paper, on the other hand, we are interested in devising
mathematically informed rewiring strategies. 
For comparison purposes, however, we will compare our rewiring methods %based on
%the updating and downdating techniques previously introduced to the random method. 
to the random rewire method, {\tt random},  
which downdates an edge (chosen uniformly at random
among all edges whose removal does not disconnect the network) and then updates a 
virtual edge, also chosen uniformly at random.

Combining the various downdating and updating methods previously introduced  
we obtain different rewiring strategies based on the centralities of edges and 
on the bounds for the total communicability. 
Concerning the methods based on the edge subgraph, eigenvector,
and total communicability centrality, 
we note that since a single downdate does not dramatically change 
the communication capability of the network, we do not need to recompute 
the centralities and the ranking of the edges after each downdating step,
at least as long as the number of rewired edges remains relatively small 
(numerical experiments not shown here support this claim).
On the other hand, after each update we may or may not recalculate
the edge centralities. As before, we use {\tt subgraph}/{\tt subgraph.no},
{\tt eigenvector}/{\tt eigenvector.no} and {\tt nodeTC}/{\tt nodeTC.no} to
refer to these three variants of rewiring.
Additionally,
we introduce another rewiring strategy, henceforth referred to
as {\tt node}, based on the subgraph centrality of the nodes. 
In this method we disconnect the most central node from the least central node among 
its immediate neighbors;
then we connect it to the most central node among those it is not linked to.
It is worth emphasizing that this strategy is philosophically different from the 
previous ones based on the edge subgraph centrality in the downdating phase 
(the updating step is the same). 
In fact, in those methods we use information on the nodes in order to 
deduce some information on the edges connecting them; on the other hand, 
the {\tt node} algorithm does not take into account the potentially 
high ``payload'' of the edges involved, whose removal may result in a 
dramatic drop in the total communicability.

\section{Numerical studies}
\label{sec:test_tc}
In this section we discuss the results of numerical studies performed in order
to assess the effectiveness and efficiency of the proposed techniques.
The tests have been performed on both synthetic and real-world networks, 
as described below. 
%We omit here for brevity the tests performed on 4 small networks to 
% assess the valuability 
%of our heuristics. 
%The description of these networks, as well as the tests and the comments 
%on the results obtained, 
%can be found in the Supplementary Material accompanying the paper. 
We refer to the Supplementary Materials for the results of computations
performed on four small social networks, aimed at comparing our heuristics with
{\tt optimal}. These results show that for these small networks, the resulting
total communicabilities are essentially identical to those obtained with the
{\tt optimal} strategy.

\subsection{Real-world networks}
\begin{table}[t]
\footnotesize
\centering
\caption{Description of the Data Set.}
\label{tab:Datasets}
\begin{tabular}{cccccc}%{|c|c|c||c|c|c|}
\hline
NAME & $n$ & $m$ &  $\lambda_1$ & $\lambda_2$ & $\lambda_1-\lambda_2$ \\
\hline
%Zachary & 34 & 78  & 6.726 & 4.977 & 1.749  \\
%Sawmill & 36 & 62  & 4.972 & 3.271 & 1.701 \\
%social3 & 32 & 80 & 5.971 & 3.810 & 2.161  \\
%dolphins & 62 & 159 & 7.193 & 5.936 & 1.257 \\
Minnesota & 2640 & 3302  & 3.2324 & 3.2319 & 0.0005  \\
USAir97 & 332 & 2126 & 41.233 & 17.308 & 23.925  \\
as--735 & 6474 & 12572 &  46.893 & 27.823 & 19.070  \\
Erd\"os02 & 5534 & 8472 & 25.842 & 12.330 & 13.512  \\
ca--HepTh & 8638 & 24806& 31.034  & 23.004  & 8.031 \\
as--22july06 & 22963  & 48436   & 71.613  & 53.166  & 18.447 \\
usroad--48     & 126146 & 161950  & 3.911    & 3.840    & 0.071 \\
\hline
\end{tabular}
\end{table}

%The real-world networks used in the tests (see Table \ref{tab:Datasets}) 
%come from a variety of sources. 
%The Zachary Karate Club network is a 
%classic example in social network analysis 
%\cite{Zachary1977}. The Sawmill and social3 networks were provided to us 
%by Prof.~Ernesto Estrada. 
%The Sawmill network describes a communication network within a 
%small enterprise (see \cite{Michael1997, Nooy2004}),

%%: 
%%all the employees were asked to indicate the frequency with which 
%%they discussed work matters with each of their colleagues in a 
%%five-point scale ranging 
%%from ``less than once a week'' to ``several times a day'' 
%%(see \cite{Michael1997, Nooy2004}).
%whereas social3 is a network of social contacts among college students 
%participating in a leadership course (see \cite{Zeleny1950}).
%%: the students choose the three 
%%members they wished to 
%%include in a committee (see \cite{Zeleny1950}).
All the %other 
networks used in the tests can be found in the University of 
Florida Sparse Matrix Collection \cite{Davis} under different ``groups''.
%The network dolphins (see \cite{Lusseau2003}) is in the Newman group 
%and represents a social network of frequent associations between 62 dolphins in a 
%community living in the waters off New Zealand.
The USAir97 and Erd\"os02 networks are from the Pajek group. 
The USAir97 network describes the US Air flight routes in 1997, while 
the Erd\"os02 network represents the Erd\"os collaboration network, 
Erd\"os included. 
The network as--735, from the SNAP group, is the communication network 
of a group of autonomous system (AS) measured over 735 days between 
November 8, 1997 and January 2, 2000. 
Communication occurs when routers from two ASs exchange information.
The Minnesota network from the Gleich group represents the Minnesota road network. 
These latter three networks are not connected, therefore the tests were 
performed on their largest connected component.
We point out that the original largest connected component 
of the network as--735 has 1323 ones on the main diagonal 
which were retained in our tests.
The network ca--HepTh is from the SNAP group and represents the 
collaboration network of arXiv High Energy Physics Theory;
the network as--22july06 is from the Newman group and represents the 
(symmetrized) structure of Internet routers as of July 22, 2006.
Finally, the network usroad--48, which is from the Gleich group, 
represents the continental US road network. For each network,
Table \ref{tab:Datasets} reports the number of nodes ($n$), 
the number of edges ($m$), the 
two largest eigenvalues, and the spectral gap.
We use the first four networks to test all methods described
in the previous section (except for {\tt optimal}, which is only
applied to the four smallest networks --- see the Supplementary Materials)
 and the last three to illustrate the
performance of the most efficient among the methods tested.

\begin{figure}[t]
\centering
\caption{Evolution of the normalized total communicability vs.~number 
of downdates, updates and rewires %% performed on large networks. 
%%10\% of nodes  selected according to the eigenvector centrality for networks 
for networks Minnesota and as735.} %%Erd\"os02. 20\% of nodes selected according 
%%to the eigenvector centrality for the network USAir97.}
\label{fig:Minnesota_as735}
\includegraphics[width=.9\textwidth]{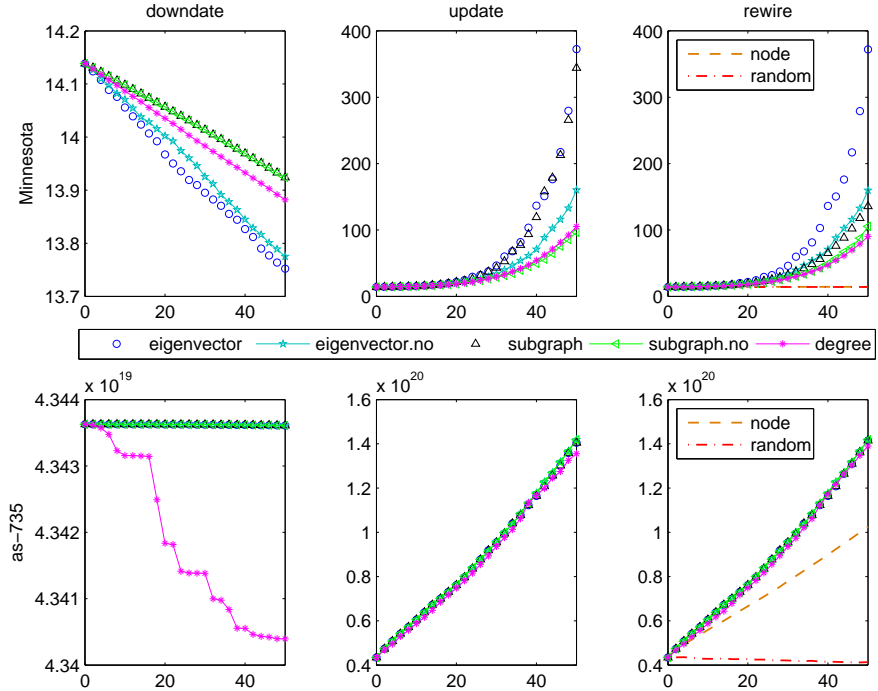}
\end{figure}

\begin{figure}[h]
\centering
\caption{Evolution of the normalized total communicability vs.~number
of downdates, updates and rewires for networks USAir97 and Erd\"os02.}
\label{fig:USAir97_Erdos02}
\includegraphics[width=.9\textwidth]{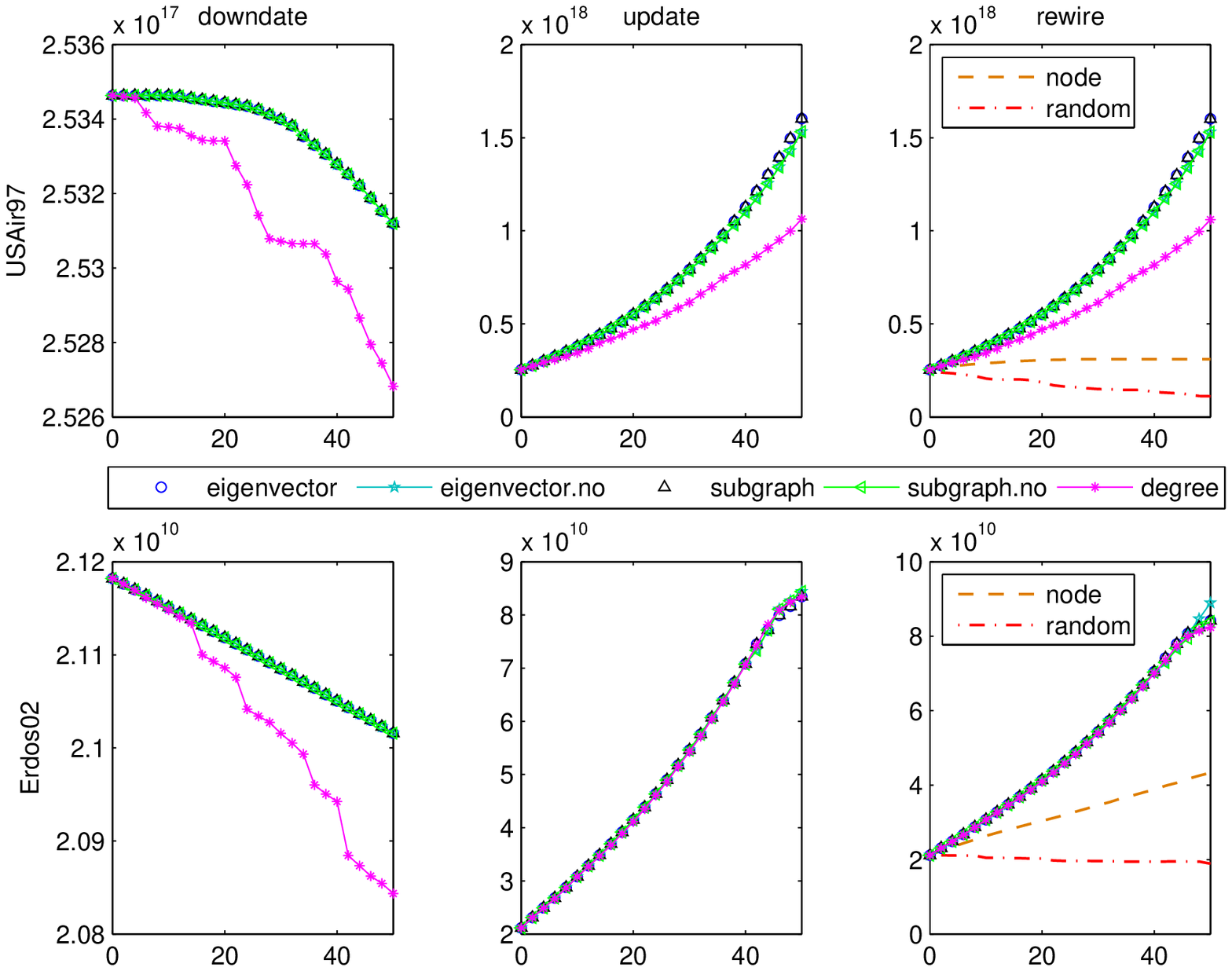}
\end{figure}

%We first consider the medium size networks 
We first consider the networks 
Minnesota, as735, USAir97, and Erd\"os02, for which we perform $K=50$ modifications. 
%%The results are shown in Figures \ref{fig:Minnesota_as735}
%%and \ref{fig:USAir97_Erdos02}.
For these networks the set $\overline{E}$ 
(the complement of the set $E$ of edges) is large enough 
that performing an extensive search for the 
edge to be updated is expensive. Hence, we form the set $S$ containing the top
$10\%$ of the nodes ordered according to the eigenvector centrality 
and we restrict our search to virtual edges incident to these nodes
only. An exception is
the network USAir97 where we have used the set $S$ corresponding to
the top $20\%$ of the nodes, since in the case of $10\%$ this set contained 
only 52 virtual edges.
In Figures \ref{fig:Minnesota_as735} and \ref{fig:USAir97_Erdos02}
we show results for the methods {\tt eigenvector}, {\tt eigenvector.no},
{\tt subgraph}, {\tt subgraph.no} and {\tt degree}.  
Before commenting on these, we want to stress the poor performance of {\tt node} 
when tackling (P3); this shows that the use of edge centrality measures (as opposed to 
node centralities alone) is indispensable in this framework. 
The results for these networks clearly show 
the effectiveness of
the {\tt eigenvector} and {\tt subgraph} algorithms
and of their less expensive variants {\tt eigenvector.no} and {\tt subgraph.no} in
nearly all cases; similar results were obtained with
{\tt nodeTC} and {\tt nodeTC.no} (not shown). The only exception is in the downdating
of the Minnesota network,
where the eigenvector-based techniques
give slightly worse results.
This fact is easily explained in view of the tiny spectral
gap characterizing this and similar networks\footnote{Small spectral gaps are typical
of large, grid-like
networks such as the road networks or the graphs corresponding to
uniform triangulations or discretizations of physical domains.}
(see Table \ref{tab:Datasets}).
Because of this property, eigenvector centrality is a poor approximation
of subgraph centrality and cannot be expected to give
results similar to those obtained with {\tt subgraph}
and {\tt subgraph.no}. 

The results for the downdate show that the inexpensive {\tt degree}
method does not perform as well on these networks, except
perhaps on Minnesota. The relatively poor performance of
this method is due to the fact that the information used by this
method to select an edge for downdating is too local.
%%Consider for instance a graph $G$ constructed as follows.
%%Start with two star graphs
%%$S_{k_1}$, $S_{k_2}$ (with $k_1,k_2$ large)
%%and connect their centers $v_1$, $v_2$ through the addition
%%of a long path graph starting at $v_1$ and ending at $v_2$;
%%Finally, add an edge to directly join one of the leaves in
%%$S_{k_1}$ with one  of the leaves in $S_{k_2}$.
%%Then, when applied to $G$, {\tt degree} will downdate
%%precisely the edge joining the two leaves (recall that
%%we are not allowed to disconnect the graph). Such a
%%choice would heavily penalize the total communicability
%%of the network, since it would eliminate the only
%%shortcut between the two star subgraphs. 
%% THIS IS WRONG!

Note, however, the scale on the vertical axis in Figures 
\ref{fig:Minnesota_as735}--\ref{fig:USAir97_Erdos02}, 
suggesting that for these networks (excluding perhaps Minnesota)
all the edge centrality-based methods perform well with only very small relative
differences between the resulting total communicabilities. 

%%In conclusion, when performing a downdate of an edge in a 
%%network it seems that the better strategy is the use of \emph{eigenvector}, but 
%%for the case of networks with small spectral gap. 
%%In this latter case the use of \emph{subgraph} is preferable.

Overall, these results indicate that the edge centrality-based
methods, especially the inexpensive {\tt eigenvector.no} and {\tt nodeTC.no} variants, 
are an excellent choice in almost all cases and to tackle all the problems. 
In the case of downdating
networks with small spectral gaps, 
{\tt subgraph.no} may be preferable but at a higher cost.

The behavior of the {\tt degree} method depends strongly on the 
network on which it is used. Our tests indicate that it behaves
well in some cases (for example, P2 for Erd\"os02) but poorly in others (P2 for Minnesota).
We speculate that this method may perform adequately when tackling (P2) on scale-free
networks (such as Erd\"os02) where a high degree is an indication
of centrality in spreading information across the network.

Some comments on the difference in the results for updating as compared to those 
for rewiring (downdating followed by updating) are in order.
Recall that our downdating strategies aim to reduce as little as 
possible the decrease in the value of the total communicability, whereas the 
updating techniques aim to increase this index as much as possible.
With this in mind, it is not surprising to see that the 
trends of the evolution of the total communicability after rewiring reflect those 
obtained with the updating strategies.
 The values obtained %%for the smaller networks using 
using the updates are in general higher than those obtained using the rewiring 
strategies, since updating implies the addition of edges whereas in
rewiring the number of edges remains the same. 
%The difference is
%especially pronounced for the small networks (except for dolphins), 
%where the effects of downdates have a greater impact,
%leading to a decrease of up nearly $70\%$ of the original value 
%of the total communicability after 25 downdates (cf.~Figure \ref{fig:down}). 
Experiments not reported here indicate that
the methods based on the edge eigenvector 
and total communicability
centrality are more stable than the others 
under rewiring and to dampen the effect of the downdates.% even for small networks. 

\begin{figure}[t]
\centering
\caption{Downdates for large networks: normalized total communicability vs.~number of modifications.}
\includegraphics[width=.9\textwidth]{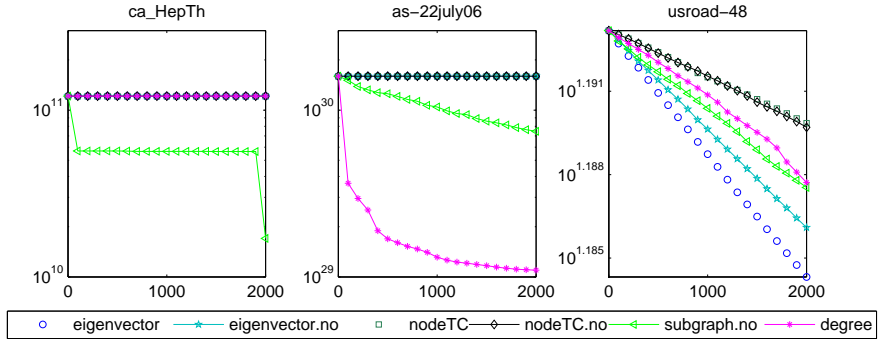}
\label{fig:large_down}
\end{figure}

\begin{figure}[t]
\centering
\caption{Updates for large networks: normalized total communicability vs.~number of modifications.}
\includegraphics[width=.9\textwidth]{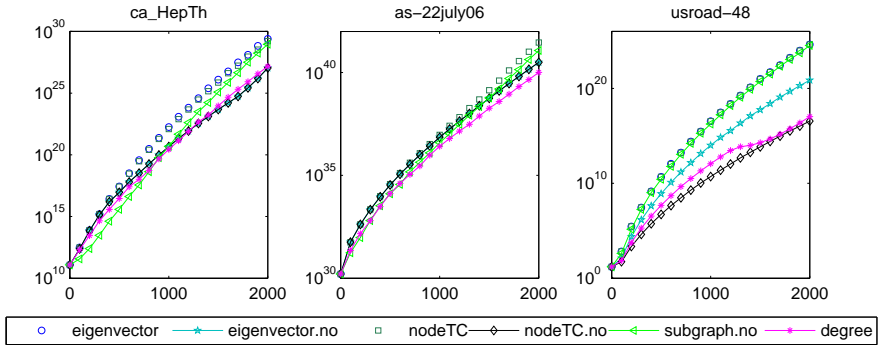}
\label{fig:large_up}
\end{figure}

%Finally, 
In Figures \ref{fig:large_down}-\ref{fig:large_up} we show results for 
the three largest networks in our data set (ca--HepTh, as-22july06 and
usroad-48). 
In the case of the updating, we have selected the virtual edges among those in the subgraph 
containing the top $1\%$ of nodes ranked according to the eigenvector centrality.  
%and the edges among them. 
We compare the following methods:
{\tt eigenvector}, {\tt eigenvector.no}, {\tt nodeTC}, {\tt nodeTC.no},
{\tt subgraph.no} and {\tt degree}; 
random downdating was also tested and found to give poor results. 
Note that network
usroad-48 behaves similarly to Minnesota; this is not surprising in view of
the fact that these are both road networks with a tiny spectral gap.
Looking at the scale on the vertical axis, however, it is clear that
the decrease in total communicability is negligible with all the methods
tested here.
The results on these networks confirm the general trend observed so far;
%%although the results obtained with {\tt subgraph.no} are not quite
%%as good as in the earlier cases.
in particular, we note the excellent behavior of {\tt nodeTC} and
{\tt nodeTC.no}.

\subsection{Synthetic networks}
The synthetic examples used in the tests were produced using 
the CONTEST toolbox for Matlab (see \cite{Contest,Taylor2009}).
We tested two types of graphs: the preferential attachment 
(Barab\'asi--Albert) model and the small world (Watts--Strogatz) model. 
%%In CONTEST these graphs and the corresponding adjacency matrices 
%%can be built using the functions \texttt{pref} and \texttt{smallw}, respectively.

\begin{figure}[t]
\centering
\caption{Evolution of the total communicability when 50 downdates,
updates or rewires are performed on two synthetic networks with $n=1000$ nodes.}
\includegraphics[width=.9\textwidth]{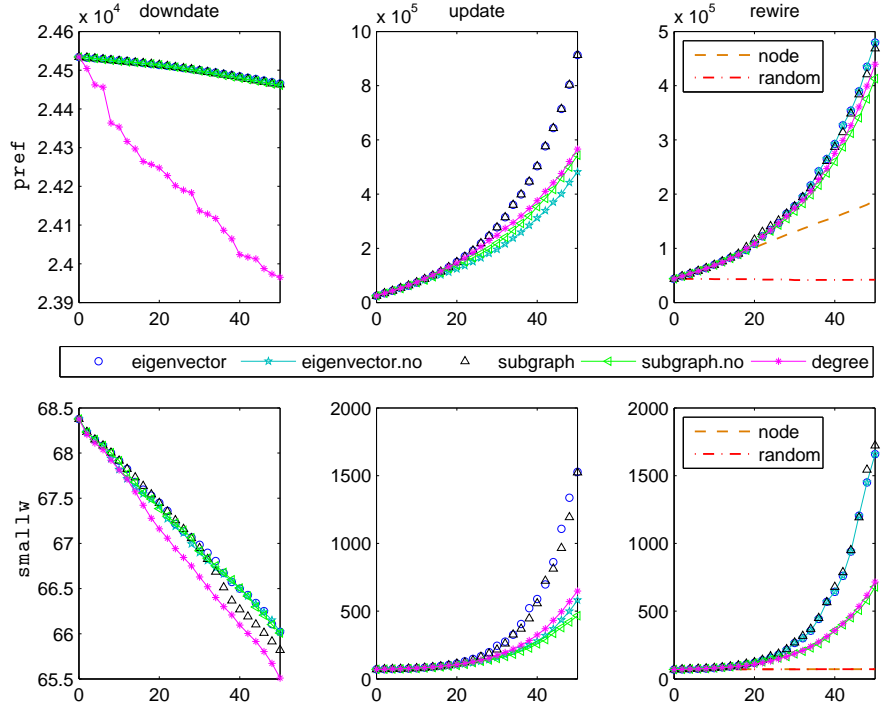}
\label{fig:synthetic}
\end{figure}

The preferential attachment model 
\cite{prefattach} was designed to produce networks with 
scale--free degree distributions as well as the small world property,
characterized by short average path length and relatively high clustering
coefficient.
In CONTEST, preferential attachment networks are constructed using 
the command \texttt{pref(n,d)} where $n$ is the number of nodes and $d\geq 1$ is the 
number of edges each new node is given when it is first introduced to the network.
The network is created by adding nodes one by one (each new node with $d$ edges).
The edges of the new node connect to nodes already in the network with 
a probability proportional to the degree of the already existing nodes.
This results in a scale--free degree distribution.

%[ {\bf CUT?} Note that with this construction, the minimum degree of the network is $d$.
%When $d>1$ this means that the network has no dangling nodes 
%(nodes of degree 1), whereas in many real-world networks one often 
%observes a high number of 
%dangling nodes.]

The second class of synthetic test matrices used in our experiments 
corresponds to Watts--Strogatz small world networks.
The small world model was developed as a way to impose a high 
clustering coefficient onto classical random graphs \cite{Watts1998}.
%%The name comes from the fact that, like classical random graphs, 
%%the Watts--Strogatz model produces networks with the small world %(that is, small graph 
%%%diameter) 
%%property.
The function used to build these matrices takes the form \texttt{smallw(n,k,p)}.
%%as described in section \ref{sec:bounds}.
Here $n$ is the number of nodes in the network, originally arranged in a
ring and connected to their $k$ nearest neighbors. Then each node is
considered independently and, with probability $p$, an edge is added
between the node and one of the other nodes in the graph, chosen uniformly
at random (self-loops and multiple edges are not allowed). 
In our tests, we have used matrices with $n=1000$ nodes which were 
built using the default values for the functions previously described. 
We used $d=2$ in the Barab\'asi--Albert model  
and $k=2$, $p=0.1$ in the Watts--Strogatz model.

The results for our tests are presented in Figure \ref{fig:synthetic}.
These results agree with what we have seen previously on real-world networks. 
Interestingly, {\tt degree} does not perform well for the downdate when 
working on the preferential attachment model; 
this behavior reflects what we have seen for the networks USAir97, as--735, 
and Erd\"os02, %% (see Fig. \ref{fig:down_big}) 
which are indeed scale--free networks.

\begin{figure}[t]
\centering
\caption{Timings in seconds for scale-free graphs of increasing size (500 modifications).} 
\includegraphics[width=.9\textwidth]{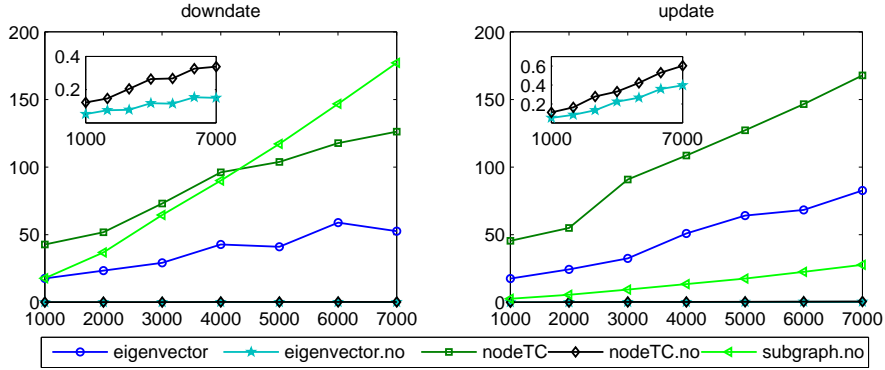}
\label{fig:timings_nest}
\end{figure}

%%%%%%%%%%%%%%%%%%%%%%%%%%%%%%%%%%%%%%%%%%%%%%%%%%%%%%%%%%%%%%%%%%%%%%%%%%%%%%%%
%%%%%%%%%%%%%%%% TIMINGS & FIGURE LARGE NETWORKS %%%%%%%%%%%%%%%%%%%%%%%%%%%%%%%
%%%%%%%%%%%%%%%%%%%%%%%%%%%%%%%%%%%%%%%%%%%%%%%%%%%%%%%%%%%%%%%%%%%%%%%%%%%%%%%%

\subsection{Timings for synthetic networks}\label{sec:ltime_synth}
We have performed some experiments with synthetic networks of increasing
size in order to assess the scalability of the various methods introduced
in this paper.  A sequence of seven adjacency matrices corresponding to
Barab\'asi--Albert scale-free graphs was generated using the CONTEST
toolbox. The order of the matrices ranges from 1000 to 7000; the average
degree is kept constant at 5. A fixed number of modifications ($K=500$)
was carried out on each network.
All experiments were performed using Matlab Version 7.12.0.635 (R2011a) 
on an IBM ThinkPad running Ubuntu 12.04.5 LTS, a 2.5 GHZ Intel Core i5 processor, and 
3.7 GiB of RAM. 
We used the built-in Matlab function {\tt eigs} (with the default settings) to
approximate the dominant eigenvector of the adjacency matrix $A$, 
the Matlab toolbox {\tt mmq} \cite{mmq} to estimate the diagonal 
entries of $e^A$ (with a fixed number of five nodes in the Gauss--Radau
quadrature rule, hence five Lanczos steps per estimate),
and the toolbox {\tt funm\_kryl} to compute the vector 
$e^A{\bf 1}$ of total communicabilities, 
also with the default parameter settings.

The results are shown in Figure \ref{fig:timings_nest}. The approximate
(asymptotic) linear scaling
behavior of the various methods (in particular of {\tt nodeTC.no} and {\tt eigenvector.no},
which are by far the fastest, see the insets) is clearly displayed 
in these plots.

\subsection{Timings for larger networks}\label{sec:large}

In Tables \ref{tab:timings_down}--\ref{tab:timings_up} we 
report the timings for various methods 
when $K=2000$ downdates and updates are selected for the three largest networks listed 
in Table \ref{tab:Datasets}. 
%%These networks can all be found in the University of Florida sparse matrix 
%%collection \cite{Davis} under different groups. 
%%The network ca--HepTh is from the SNAP group and represents the collaboration 
%%network of arXiv High Energy Physics Theory;
%%the network as--22july06 is from the Newman group and represent the (symmetrized) 
%%structure of Internet routers as of July 22, 2006. 
%%Finally, the network usroad--48, which is from the Gleich group, represents the 
%%continental US road network.

The timings presented refer to the selection of the edges to be downdated or updated, which
dominates the computational effort. For the method {\tt subgraph.no} in the case
of downdates, we restricted the search of
candidate edges to a subset of $E$ in order to reduce
costs. For the three test networks we used $40\%$, $45\%$ and $15\%$ of the nodes,
respectively,
chosen by taking those with lowest eigenvector centrality, and the corresponding
edges. We found the results to be very close to those obtained working with the
complete set $E$, but at a significantly lower cost (especially for the largest
network).

These results clearly show that algorithms
{\tt nodeTC.no} and {\tt eigenvector.no} are orders of magnitude
faster than the other methods; method {\tt subgraph.no}, while significantly
more expensive, is still
reasonably 
efficient\footnote{It is worth mentioning that in principle it is possible to 
greatly reduce the cost of this method using parallel processing, since each 
subgraph centrality can be computed independently of the others.}
and can be expected to give better results in 
some cases (e.g.,
on networks with a very small spectral gap). The {\tt degree} algorithm, on the
other hand, cannot be recommended in general since it gives somewhat inferior results.
The remaining methods {\tt eigenvector}, {\tt nodeTC} and {\tt subgraph} (not
shown here) are prohibitively expensive for large networks, at least when the
number $K$ of modifications is high (as it is here).

\begin{table}[t]
\footnotesize
\centering
\caption{Timings in seconds for $K=2000$ downdates performed on the 
three largest networks in Table \ref{tab:Datasets}.}
\label{tab:timings_down}
\begin{tabular}{cccc}
\hline
                         & ca--HepTh & as--22july06 & usroad--48  \\
\hline
{\tt eigenvector}   & 278.13     & 599.83        & 11207.39  \\
%%{\tt eigenvector}   & 270.06     & 569.93        & 11239.78  \\
{\tt eigenvector.no}& 0.07       & 1.79          & 4.08      \\
%%{\tt eigenvector.no}& 0.07       & 0.16          & 3.39      \\
{\tt nodeTC}        & 553.04     & 1234.49       & 2634.27   \\
%%{\tt nodeTC}        & 539.90     & 1192.35       & 2658.49   \\
{\tt nodeTC.no}     & 0.34       &   0.83        &   1.34    \\
%%{\tt nodeTC.no}     & 0.38       &   0.65        &   1.34    \\
%%{\tt subgraph.no}   & 246.44     & 808.24        & 10322.41  \\
{\tt subgraph.no}   & 107.36     & 383.34        & 1774.07  \\
{\tt degree}         & 29.67      & 53.42         & 153.52    \\
%%{\tt degree}         & 21.23      & 47.60         & 156.39    \\
\hline
\end{tabular}
\end{table}

\begin{table}[t]
\footnotesize
\centering
\caption{Timings in seconds for $K=2000$ updates performed on the 
three largest networks in Table \ref{tab:Datasets}}
\label{tab:timings_up}
\begin{tabular}{cccc}
\hline
                         & ca--HepTh & as--22july06 & usroad--48\\
\hline
{\tt eigenvector}   & 192.8     & 436.9        & 1599.5    \\
{\tt eigenvector.no}& 0.19      & 0.33         & 5.85      \\
{\tt nodeTC}        & 561.9    & 1218.8      & 2932.   \\
{\tt nodeTC.no}     & 0.30      &  0.55        &  1.59     \\
{\tt subgraph.no}   & 3.13      & 7.20         & 121.4     \\
{\tt degree}         & 11.1      & 12.4         & 175.8     \\
\hline
\end{tabular}
\end{table}

We also observe that downdating is generally a more
expensive process than updating, since in the latter case the edges are to be
chosen among a fairly small subset of all virtual edges, whereas in the downdating process
we work on the whole set $E$ of existing edges (or on a large subset of $E$). 
For some methods the difference in cost becomes significant when
the networks are sufficiently large and the number of modifications to be
performed is high.

Summarizing, the method labelled {\tt nodeTC.no} is the fastest and gives excellent
results, quite close to those of the more expensive methods, and therefore we can
recommend its use for the type of problems considered here. The methods labelled
{\tt eigenvector.no} and {\tt subgraph.no} are also effective and
may prove useful in some settings, especially for updating.

%%%%%%%%%%%%%%%%%%%%%%%%%%%%%%%%%%%%%%%%%%%%%%%%%%%%%%%%%%%%%%%%%%%%%%%%%%
%%%%%%%%%%%%%%%%%%% EVOLUTION OF OTHER PROPERTIES %%%%%%%%%%%%%%%%%%%%%%%%
%%%%%%%%%%%%%%%%%%%%%%%%%%%%%%%%%%%%%%%%%%%%%%%%%%%%%%%%%%%%%%%%%%%%%%%%%% 

\section{Evolution of other connectivity measures}
\label{sec:nc+gap}
In this section we want to highlight another facet of the methods we 
have introduced for (approximately) optimizing the total communicability.
In particular, we
look at the evolution of other network properties under our updating strategies.
When building or modifying a network, there are various 
features that one may want to achieve.
Typically, there are two main desirable properties: first, the network should
do a good job at spreading information, i.e., have a high total communicability;
second, the network should be robust under targeted attacks or random failure, which 
is equivalent to the requirement that it should be difficult to ``isolate'' 
parts of the network, i.e., the network should be ``well connected''.
This latter property can be measured by means of various indices. 
One such measure is the spectral gap $\lambda_1 - \lambda_2$. As a consequence
of the Perron--Frobenius Theorem, adding an edge to a connected network
causes the dominant eigenvalue $\lambda_1$ of $A$ to increase. Test results
(not shown here) show that when a network is updated using one of our
techniques, the first eigenvalue increases rapidly with the number of updates.
On the other hand, the second eigenvalue $\lambda_2$ tends to change little
with each update and it may even decrease (recall that the matrix
$UW^T = {\bf e}_i{\bf e}_j^T + {\bf e}_j{\bf e}_i^T$ being added to $A$
in an update is indefinite). Therefore, the spectral gap $\lambda_1 - \lambda_2$
widens rapidly with the number of updates.\footnote{This fact, incidentally,
may serve as further justification for the effectiveness
of algorithms like {\tt nodeTC.no}
and {\tt eigenvector.no}.}
 It has been pointed out
by various
 authors (see, e.g., \cite{E06b,Puder2013}) that a large spectral gap is typical
of complex networks with good expansion properties.  

Here we focus on a related measure, the so-called {\em free energy}
(also known in the literature as {\em natural connectivity}) of the network.
In particular, we investigate the effect of our proposed methods of network updating
on the evolution of this index.

\subsection{Tracking the free energy (or natural connectivity)}
In \cite{Jun2010} the authors discuss a measure of network connectivity which 
is based on an intuitive notion of robustness and
whose analytical expression has a clear meaning and can be derived from the 
eigenvalues of $A$; they refer to it as the {\em natural connectivity}
of the network (see also \cite{Wu2012}).
The idea underlying this index is that a network is more robust if there exists 
more than one route to get from one node to another; this property 
ensures that if a route becomes unusable, there is an alternative way to get 
from the source of information to the target.
Therefore, intuitively a network is more robust if it has a lot of (apparently) 
redundant routes connecting its vertices  
or, equivalently, if each of its nodes is involved in a lot of closed walks.
The natural connectivity aims at quantifying
this property by using an existing measure for
the total number of closed walks in a graph, namely, the 
{\em Estrada index} \cite{Estrada2000}.
This index, denoted by $EE(G)$, 
is defined as the trace of the matrix exponential.
%Recall that the Estrada index of a graph is 
%defined as
%$$EE(G)=\sum_{i=1}^ne^{\lambda_i}.$$ 
Normalizing this value and taking the natural logarithm, 
one obtains the {\itshape natural connectivity} 
%(or {\itshape natural eigenvalue}) 
of the graph:
$$\overline{\lambda}(A)=\ln\left(\frac{1}{n}\sum_{i=1}^ne^{\lambda_i}\right)=\ln(EE(G))-\ln(n).$$ %\ln\left(\frac{EE(G)}{n}\right).$$

%which can be seen as an ``average'' eigenvalue and changes monotonically 
%when an edge is downdated or updated in the graph (see \cite{Jun2010}).

%Coarse bounds on this index are readily obtained:
%$$0\leq\overline{\lambda}(A)\leq\ln((n-1)e^{-1}+e^{n-1}) - \ln n.$$
%The lower bound is attained by the empty graph, while the upper bound is attained by 
%the complete graph, as a straightforward computation shows.
%Using the results in \cite{Benzi2010} we obtain more refined bounds via quadrature rules:
%$$\ln\left(\frac{1}{n}\sum_{i=1}^n\frac{\beta^2e^{\frac{d_i}{\beta}}+
%d_ie^{-\beta}}{\beta^2+d_i}\right)\leq \overline{\lambda}(A) \leq
%\ln\left(\frac{1}{n}\sum_{i=1}^n\frac{\alpha^2e^{\frac{d_i}{\alpha}}+
%d_ie^{-\alpha}}{\alpha^2+d_i}\right),$$
%where $[\alpha,\beta]$ is an interval containing the spectrum of $-A$ and $d_i$ is 
%the degree of node $i$.

It turns out, however, that essentially the same index was already
present in the literature. Indeed, the natural connectivity is only one 
of the possible interpretations
that can be given to the logarithm of the (normalized) Estrada index.
Another, earlier interpretation was given in \cite{Estrada2007}, where the authors
related this quantity to the Helmholtz free energy of the network $F=-\ln\left(EE(G)\right)$.
Therefore, since $\overline{\lambda}=-F-\ln(n)$, the behavior of $F$ completely describes that of
$\overline{\lambda}$ (and conversely) as the graph is modified by adding or removing links.

%%In particular, since the evolution of the natural connectivity has
%%been shown to mirror the evolution of the total communicability,
%%from this latter we can infer the change in the free energy of the network.

The natural connectivity has been recently used (see \cite{Chan2014}) to derive manipulation 
algorithms that directly optimize this robustness measure. 
In particular, the updating algorithm introduced in \cite{Chan2014} appears to be 
superior to existing heuristics, such as those proposed in 
\cite{Beygelzimer2005,Frank1970,Shargel2003}. 
%%the strategies based on the idea that similar 
%%nodes tend to connect with each other (rich with rich and poor with poor, 
%%where a node is considered rich if it has a high degree, poor otherwise) 
%%and with the updating methods based on the counterintuitive idea that rich and 
%%poor nodes may connect.
This algorithm, which costs $O(mt+Kd_{max}^2t+Knt^2)$ where $d_{max}=\max_{i\in V}d_i$ 
and $t$ is the (user-defined) number of leading eigenpairs, 
selects $K$ edges to be added to the network % as 
%described in Algorithm \ref{algo:Chan}. 
%%We have compared our updating techniques with that described in Alg. \ref{algo:Chan}.
by maximizing a quantity that involves the elements of the leading $t$ 
eigenpairs of $A$.\footnote{A description of the 
algorithm can be found in the Supplementary Material.}

%\begin{algorithm}[t]
%\footnotesize
%\caption{\footnotesize{Updating algorithm from \cite{Chan2014}.} }
%\label{algo:Chan}
%%\SetLine 
%% \SetAlgoLined
% \KwData{$A$ adjacency matrix and $K\in\mathbb{N}$}
% \KwResult{Set $\mathcal{S}$ of $K$ edges to be added}
%$\mathcal{S}=\emptyset$\; 
%Compute the top $t$ eigenpairs $(\lambda_k,\mathbf{q}_k)$ of $A$\;
%\For{iter = 1 : $K$}{
%  Compute $d_{\max}=\max(d_i)$, the largest degree of $A$ \;
%  Find the set $C$ of $d_{\max}$ nodes with the highest eigenvector centrality\;
%  Select the edge $(i^*,j^*)\in\overline{E}$  that maximizes 
%  $$
%  e^{\lambda_1} \left( e^{2q_1(i)q_1(j)} + \sum_{h=2}^te^{\lambda_h-\lambda_1}e^{2q_h(i)q_h(j)}\right)
%  $$
%  and such that $i^*,j^*\in C$, $i^*\neq j^*$\;
%  $\mathcal{S}=\mathcal{S}\cup \{(i^*,j^*)\}$, $E=E\cup \{(i^*,j^*)\}$\;
%  Update $A$\;
%  Update the top $t$ eigenpairs as
%$$\left\{
%\begin{array}{l}
%\lambda_k=\lambda_k+2q_k(i)q_k(j); \\
%\mathbf{q}_k=\mathbf{q}_k+\sum_{h\neq k}\left(\frac{q_h(i)q_k(j)-q_k(i)q_h(j)}{\lambda_k-\lambda_h}\mathbf{q}_h\right)
%\end{array}\right.\qquad k=1,2,\ldots,t;$$
%}
%Return $\mathcal{S}$.
%\end{algorithm}

\begin{figure}[t]
\caption{Evolution of the natural connectivity and of the normalized total 
communicability (in a semi--logarithmic scale plot) when up to 500 updates are 
performed on four real-world networks.}
\centering
\includegraphics[width=.9\textwidth]{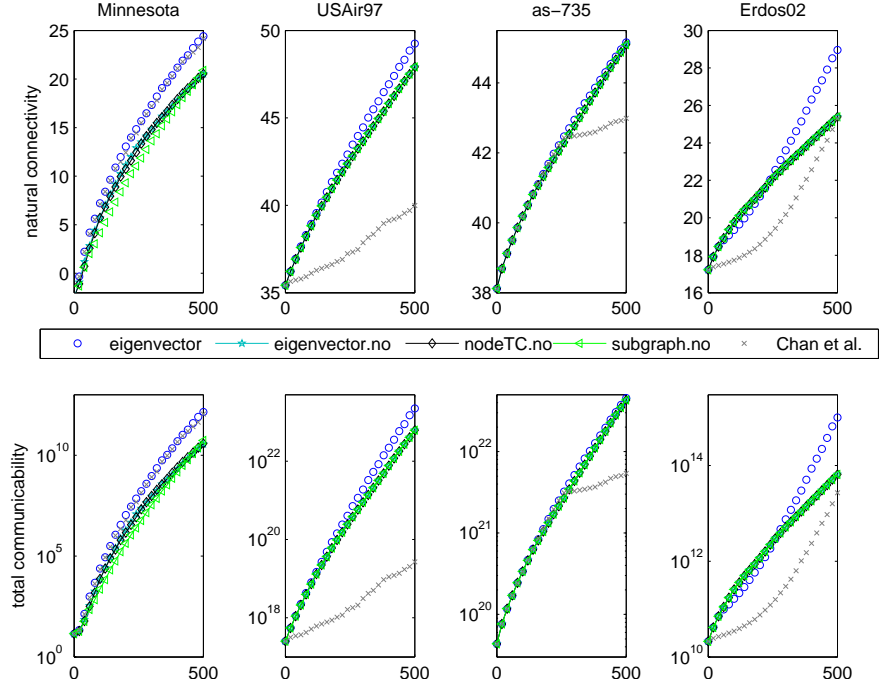}
\label{fig:TCeNC_big}
\end{figure}

We have compared our updating techniques with that described in \cite{Chan2014}.
Results for four representative networks are shown in Figure \ref{fig:TCeNC_big}.
In our tests, we use the value $t=50$ (as in \cite{Chan2014}), and we select $K=500$ edges. 
Note that, when $K$ is large, the authors recommend to recompute the set of $t$ 
leading eigenpairs every $l$ iterations. 
This operation requires an additional effort that our faster methods do not need.
Since the authors in \cite{Chan2014} show numerical experiments in which the methods 
with and without the recomputation return 
almost exactly the same results, we did not recompute the eigenpairs after 
any of the updates. 

Figure \ref{fig:TCeNC_big} displays the results for both the evolution of the natural 
connectivity and of the normalized total communicability, where
the latter is plotted in a semi--logarithmic scale.
A total of 500 updates have been performed.
The method labelled {\tt Chan} selects the edges according to
the algorithm described in \cite{Chan2014} choosing from all the virtual edges of the graph. 
For our methods we used, as before, the virtual edges in the subgraph obtained 
selecting the top $10\%$ or $20\%$ of nodes ranked according 
to the eigenvector centrality.
As one can easily see, our methods generally outperform the algorithm proposed in 
\cite{Chan2014}. In particular, {\tt nodeTC.no} and {\tt eigenvector.no} 
give generally better results than {\tt Chan} and
are much faster in practice. For instance, the execution time
with {\tt Chan} on the network ca-HepTh was over 531 seconds, and 
much higher for the two larger networks.
We recall (see Table \ref{tab:timings_up})
 that the execution times for {\tt nodeTC.no} and {\tt eigenvector.no}
are about three orders of magnitude smaller.

It is striking to see how closely the evolution of the natural connectivity 
mirrors the behavior of the normalized total communicability. This is likely
%%due to the fact that both measures receive a large contribution from the
%%term $e^{\lambda_1} (or from terms containing it, see (\ref{tc_spec})), and all
%%these updating strategies tend to increase $\lambda_1$.
due to the fact that both indices depend on the eigenvalues of $A$
(with a large contribution coming from the terms containing $\lambda_1$,
see (\ref{tc_spec}) and the subsequent remark), and all the updating strategies used here
tend to make $\lambda_1$ appreciably larger.

Returning to the interpretation in terms of statistical physics,  
from Figure \ref{fig:TCeNC_big} we deduce that the free energy of the
graph decreases as we add edges to the network.
In particular this means that the network is evolving toward a more stable
configuration and, in the limit, toward equilibrium, which is the configuration
with maximum entropy.\footnote{The relation between the free energy and the 
Gibbs entropy is described in more detail in the Supplementary Material.}

\begin{comment}
Indeed, the free energy of the system is related to the Gibbs entropy $S$ as
$TS=H-F$, 
where $T$ is the absolute temperature and $H$ is the total energy of the graph.
Therefore, the Gibbs entropy,
which measures the effective number of states sharing the same energy,
increases as $F$ decreases.
\end{comment}

These findings indicate
that the normalized total communicability is equally effective an index as
the natural connectivity
(equivalently, the free energy) for the purpose of 
characterizing network connectivity.
Since the network total communicability can be computed very fast (in $O(n)$ time),
we believe that the normalized total communicability should be used instead
of the natural connectivity, especially for large networks. 

Indeed, computing the natural connectivity requires evaluating the trace of 
$e^A$; even when stochastic trace estimation is used \cite{AT11}, this is 
several times more expensive, for large networks, than the total communicability.

\begin{comment}
Indeed, computing
the natural connectivity requires evaluating all the diagonal entries of $e^A$ and is
therefore significantly more expensive, for large networks, than the total
communicability.
\end{comment}

%%%%%%%%%%%%%%%%%%%%%%%%%%%%%%%%%%%%%%%%%%%%%%%%%%%%%%%%%%%%%%%%%%%%%%%%%%
%%%%%%%%%%%%%%%%%%%%% CONCLUSIONS & OUTLOOK %%%%%%%%%%%%%%%%%%%%%%%%%%%%%%
%%%%%%%%%%%%%%%%%%%%%%%%%%%%%%%%%%%%%%%%%%%%%%%%%%%%%%%%%%%%%%%%%%%%%%%%%%
\section{Conclusions and future work}
\label{sec:conclusions}
In this paper we have introduced several techniques that can be used 
to modify an existing network so as to obtain networks that are highly 
sparse, and yet have 
a large total communicability.

These heuristics make use of various measures of edge 
centrality, a few of which have been introduced in this work. 
Far from being {\em ad hoc}, these heuristics are widely
applicable and mathematically justified.
All our techniques can be implemented using well-established tools from numerical
linear algebra: algorithms for eigenvector computation, Gauss-based quadrature
rules for estimating quadratic forms, and Krylov subspace methods for computing
the action of a matrix function on a vector. At bottom, the Lanczos algorithm
is the main player. High quality, public domain software exists to perform these
modifications efficiently.

Among all the methods introduced here, the best results are obtained by 
the {\tt nodeTC.no} and {\tt eigenvector.no} algorithms, 
which are based on the edge total communicability
and eigenvector centrality, respectively. These methods are extremely fast 
and returned excellent results in virtually all 
the tests we have performed. For updating networks characterized by a small
spectral gap, a viable alternative is the algorithm {\tt subgraph.no}.
While more expensive than {\tt nodeTC.no} and {\tt eigenvector.no}, this
method scales linearly with the number of nodes and yields consistently
good results.

Finally, we have shown that the total communicability can be effectively 
used as a measure of network connectivity, which plays an important role
in designing robust networks.
 Indeed, the total communicability does a very good job at quantifying  
two related properties of networks: 
the ease of spreading information, and the extent to which the network is 
``well connected''. Our results show that the total communicability
behaves in a manner very similar to the natural connectivity (or free
energy) under network
modifications, while it can be computed much more quickly.

Future work should include the extension of these techniques
to other types of networks,
including directed and weighted ones.

%%%%%%%%%%%%%%%%%%%%%%%%%%%%%%%%%%%%%%%%%%%%%%%%%%%%%%%%%%%%%%%%%%%%%%%%%
%%%%%%%%%%%%%%%%% ACKNOWLEDGEMENTS & FUNDING %%%%%%%%%%%%%%%%%%%%%%%%%%%%
%%%%%%%%%%%%%%%%%%%%%%%%%%%%%%%%%%%%%%%%%%%%%%%%%%%%%%%%%%%%%%%%%%%%%%%%%

\section*{Acknowledgements}
We are grateful to Ernesto Estrada for providing some of the networks 
used in the numerical experiments and for pointing out some useful references.
The first author would like to thank Emory University for the hospitality 
offered in 2014, when this work was completed.
We also thank two anonymous referees for helpful suggestions.

%%%%%%%%%%%%%%%%%%%%%%%%%%%%%%%%%%%%%%%%%%%%%%%%%%%%%%%%%%%%%%%%%%%%%%%%%%
%%%%%%%%%%%%%%%%%%%%%%%%%% BIBLIOGRAPHY %%%%%%%%%%%%%%%%%%%%%%%%%%%%%%%%%%
%%%%%%%%%%%%%%%%%%%%%%%%%%%%%%%%%%%%%%%%%%%%%%%%%%%%%%%%%%%%%%%%%%%%%%%%%%

\newpage

%%%%%%%%%%%%%%%%%%%%%%%%%%%%%%%%%%%%%%%%%%%%%%%%%%%%%%%%%%%%%%%%%%%%%%%%%%
%%%%%%%%%%%%%%%%%%%% SUPPLEMENTARY MATERIAL %%%%%%%%%%%%%%%%%%%%%%%%%%%%%%
%%%%%%%%%%%%%%%%%%%%%%%%%%%%%%%%%%%%%%%%%%%%%%%%%%%%%%%%%%%%%%%%%%%%%%%%%%
\appendix

\section{Supplementary Materials to the paper}
\setcounter{figure}{0} \renewcommand{\thefigure}{A.\arabic{figure}} 
\setcounter{table}{0}\renewcommand{\thetable}{A.\arabic{table}}

\begin{abstract}
In this document we summarize a few supplementary results to the accompanying 
paper. 
We give a detailed proof of the bounds 
on the normalized total communicability obtained via quadrature rules
(Theorem \ref{thm:bounds}, section \ref{sec:bounds}). 
We also derive the computational costs for the heuristics introduced. 
Moreover, this document contains the results of some numerical experiments performed 
on four small networks to assess the valuability of our techniques. 
We provide full descriptions four our downdating and updating algorithms and for the
updating algorithm developed in \cite{A:Chan2014}. 
Finally, we briefly recall the approach used %in \cite{A:Estrada2007} 
to relate the Estrada index of a graph to its Helmholtz free energy.
\end{abstract}

\subsection{Bounds via quadrature rules: a proof of Theorem \ref{thm:bounds}}
In this section we give a proof of Theorem \ref{thm:bounds}.  
In order to make this document self-contained, we briefly recall here the 
technique used to derive bounds via quadrature rules on bilinear forms. 

Bounds on bilinear forms $\mathbf{u}^Tf(A)\mathbf{v}$ can be 
derived based on Gauss--type quadrature rules when $f$ is a 
{\itshape strictly completely 
monotonic} (s.c.m.) function on the interval $[a,b]$ containing the 
spectrum of $A$ by working on a $2\times 2$ matrix derived from one 
step of the symmetric 
Lanczos iteration (see \cite{A:Benzi1999,A:Golub2010}).
Recall that a function is s.c.m.~on $[a,b]$ if $f^{(2l)}(x)>0$ and 
$f^{(2l+1)}(x)<0$ for all $x\in [a,b]$ and for all $l\geq 0$, where $f^{(k)}$ denotes the 
$k$th derivative of $f$ and $f^{(0)}\equiv f$.
In order to compute bounds for the normalized total communicability, 
this means that we need to use 
$f(x)=e^{-x}$ and therefore we work with the matrix $-A$.

The starting point is to observe that bilinear forms 
$\mathbf{u}^Tf(A)\mathbf{v}$ can be thought of as Riemann--Stieltjes integrals  
with respect to the spectral measure associated with the symmetric matrix $A$: 
$$\mathbf{u}^Tf(A)\mathbf{v}=\int_a^bf(\lambda)dm(\lambda),\quad 
m(\lambda)=\left\{
\begin{array}{ll}
0, & \lambda<a=\lambda_n \\
\sum_{k=i+1}^nw_kz_k, & \lambda_{i+1}\leq\lambda<\lambda_i\\
\sum_{k=1}^nw_kz_k, & b=\lambda_1\leq\lambda
\end{array}\right.$$
where $A=Q\Lambda Q^T$, $\mathbf{w}=Q^T\mathbf{u}=\left(w_i\right)$, 
and $\mathbf{z}=Q^T\mathbf{v}=\left(z_i\right)$.

This integral can be approximated by means of Gauss--type quadrature 
rules, which can be used to obtain lower and upper bounds on the bilinear forms of 
interest. 
In particular, our bounds are derived using the Gauss--Radau quadrature rule:
\begin{equation}\label{eq:quadrature}
\int_a^bf(\lambda)dm(\lambda)=\sum_{j=1}^pc_jf(t_j)+v_1f(\tau_1),
\end{equation}
where the nodes $\{t_j\}_{j=1}^p$ and the weights 
$\left\{\{c_j\}_{j=1}^p, v_1\right\}$ are to be determined, whereas $\tau_1$ is prescribed 
and equal either to $a$ or to $b$.
The Gauss--Radau bounds are then as described in the following theorem.
\begin{theorem}[6.4 in \cite{A:Golub2010}]\label{lemma:bound}
Suppose $f$ is such that $f^{(2l+1)}(\xi)<0$ for all $l$ and for all $\xi\in (a,b)$. 
Let $U_{GR}$ be defined as 
$$U_{GR}[f]=\sum_{j=1}^pc_j^af(t_j^a)+v_1^af(a),$$
$c_j^a,v_1^a,t_j^a$ being the weights and nodes in (\ref{eq:quadrature}) with $\tau_1=a$, 
and let $L_{GR}$ be defined as
$$L_{GR}[f]=\sum_{j=1}^pc_j^bf(t_j^b)+v_1^bf(b),$$
$c_j^b,v_1^b,t_j^b$ being the weights and nodes in 
(\ref{eq:quadrature}) with $\tau_1=b$.
The Gauss--Radau rule is exact for polynomials of degree 
less than or equal to $2p$ and satisfies 
$$L_{GR}[f]\leq \int_{a}^bf(\lambda)dm(\lambda)\leq U_{GR}[f].$$
Moreover for all $p$ there exists $\eta_U,\eta_L\in[a,b]$ such that
$$\int_a^bf(\lambda)dm(\lambda)-U_{GR}[f]=
\frac{f^{(2p+1)}(\eta_U)}{(2p+1)!}\int_a^b(\lambda-a)
\left[\prod_{j=1}^p(\lambda-t_j^a)\right]^2dm(\lambda),$$
$$\int_a^bf(\lambda)dm(\lambda)-L_{GR}[f]=\frac{f^{(2p+1)}(\eta_L)}
{(2p+1)!}\int_a^b(\lambda-b)\left[\prod_{j=1}^p(\lambda-t_j^b)\right]^2d
m(\lambda).$$
\end{theorem}

It is therefore necessary to evaluate two quadrature rules, one for the 
upper bound and one for the lower bound.
However, the explicit computation of nodes and weights can be avoided.
Indeed, the evaluation of the quadrature rules is 
mathematically equivalent to the computation of orthogonal polynomials 
via a three--term recurrence, or, 
equivalently, to the computation of entries and spectral information 
of a certain tridiagonal matrix via the Lanczos algorithm. 
In fact, the right hand side of equation (\ref{eq:quadrature}) 
can be computed from the 
relation (Theorem 6.6 in \cite{A:Golub2010}):
\begin{equation}\label{relation}
\sum_{j=1}^{p}c_jf(t_j)+v_1f(\tau_1)=\mathbf{e}_1^Tf(J_{p+1})\mathbf{e}_1,
\end{equation}
where 
$$J_{p+1}=\left(
\begin{array}{ccccc}
\omega_1 & \gamma_1 & & & \\
\gamma_1 & \omega_2 & \gamma_2 & & \\
         & \ddots  & \ddots   & \ddots &  \\
         &         &  \gamma_{p-1} & \omega_{p} & \gamma_{p} \\
         &         &              & \gamma_{p} & \omega_{p+1} \\
\end{array}
\right)$$
is a tridiagonal matrix whose eigenvalues are the Gauss--Radau nodes 
(and hence $J_{p+1}$ is built so as to have the prescribed eigenvalue $\tau_1$), 
whereas the weights are given by the squares of the first entry of the normalized 
eigenvectors of $J_{p+1}$. An efficient implementation of this technique
is provided in G.~Meurant's {\tt mmq} toolbox for Matlab \cite{A:mmq}. This
toolbox, adapted to handle sparsity, has been used for some of the numerical experiments
presented in the paper.

We can now prove Theorem \ref{thm:bounds}, which contains the bounds for the normalized total network communicability. 
The proofs of the subsequent corollaries follow the same line.

%\begin{theorem}
%Let $A$ be the adjacency matrix of an unweighted and undirected network. Then
%\begin{equation*}
%\Phi\left(\beta,\omega_1+\frac{\gamma_1^2}{\omega_1-\beta}\right)\leq 
%\frac{TC(A)}{n}\leq\Phi\left(\alpha,\omega_1+\frac{\gamma_1^2}{\omega_1-\alpha}\right)
%\end{equation*}
%where $[\alpha,\beta]$ is an interval containing the spectrum of $-A$ 
%(i.e., $\alpha \le -\lambda_1$ and $\beta\ge -\lambda_n$), 
%$\omega_1=-\mu=-\frac{1}{n}\sum_{i=1}^nd_i$ is the negative mean of the degrees, 
%$\gamma_1=\sigma=\sqrt{\frac{1}{n}\sum_{k=1}^n(d_k-\mu)^2}$ is the standard deviation, and
%\begin{equation}\label{eq:bound}
%\Phi(x,y)=\frac{c \left(e^{-x}-e^{-y}\right)+xe^{-y}-ye^{-x}}{x-y}, \qquad c=\omega_1. 
%\end{equation}
%\end{theorem}

\begin{proof}%[of Theorem \ref{thm:bounds}]
First we derive an explicit expression for the right--hand side of equation 
(\ref{relation}) when $f(x)=e^{-x}$ and $J_2$ is $2\times 2$  
with the help of the Lagrange interpolation formula for the evaluation of matrix 
functions \cite{A:Higham2008}. 
Let $\mu_1$ and $\mu_2$ be distinct eigenvalues of a given 
$2\times 2$ matrix $B=(b_{ij})$, then
$$e^{-B}=\frac{e^{-\mu_1}}{\mu_1-\mu_2}(B-\mu_2I)+
\frac{e^{-\mu_2}}{\mu_2-\mu_1}(B-\mu_1I)$$
where $I$ is the $2\times 2$ identity matrix.
It follows that 
$$\mathbf{e}_1^T\left(e^{-B}\right)\mathbf{e}_1=\frac{b_{11}(e^{-\mu_1}-e^{-\mu_2})+
\mu_1e^{-\mu_2}-\mu_2e^{\mu_1}}{\mu_1-\mu_2}.$$

Next, we build explicitly the matrix $J_2$ and compute its eigenvalues. 
%This, together with the above description of $\left(e^{-B}\right)_{11}$
% and Theorem \ref{lemma:bound}, concludes the proof.
The values of $\omega_1=-\mu$ and $\gamma_1=\sigma$ are derived
applying one step of Lanczos iteration to the matrix $-A$ with starting vectors 
$\mathbf{x}_{-1}=\mathbf{0}$ and $\mathbf{x}_0=\frac{1}{\sqrt{n}}\mathbf{1}$. 
We want to compute the value of $\omega_2$ in such a way that the matrix
$J_2$ has the prescribed eigenvalue $\tau_1=\alpha$ or $\tau_1=\beta$. 
Note that $\gamma_1=0$ if and only if the graph is regular, i.e., if and only if 
the nodes in the graph have all the same degree.
In such case we simply 
take $\omega_2=\tau_1$ and the matrix $J_2$ is diagonal 
with eigenvalues $\mu_1=-\mu$ and $\mu_2=\tau_1$.
Thus, let us assume $\gamma_1\neq 0$. 
In order to compute the value for $\omega_2$, we use the three--term 
recurrence for orthogonal polynomials:
$$\gamma_jp_j(\lambda)=(\lambda-\omega_j)p_{j-1}(\lambda)-
\gamma_{j-1}p_{j-2}(\lambda),\quad j=1,2,\ldots,p,$$
with
$p_{-1}(\lambda)\equiv 0$, $p_0(\lambda)\equiv 1$
to impose that $p_2(\tau_1)=0$ and hence derive 
$\omega_2=\tau_1-\frac{\gamma_1}{p_1(\tau_1)}$.
Using the same recurrence, we also find that $p_1(\tau_1)=
\frac{\tau_1-\omega_1}{\gamma_1}$ which is nonzero, since the zeros of 
orthogonal polynomials satisfying the three--term recurrence are distinct and lie
in the interior of $[\alpha,\beta]$ (see \cite[Theorem 2.14]{A:Golub2010}).

Finally, the matrix
$$J_2=\left(
\begin{array}{cc}
\omega_1 & \gamma_1 \\
\gamma_1 & \tau_1-\frac{\gamma_1^2}{\tau_1-\omega_1}
\end{array}
\right)$$
has (distinct) eigenvalues $\mu_1=\tau_1$ and 
$\mu_2=\omega_1+\frac{\gamma_1^2}{\omega_1-\tau_1}$.
This, together with 
Theorem \ref{lemma:bound} and the relation (\ref{relation}), concludes the proof.
\qquad\end{proof}

%%%%%%%%%%%%%%%%%%%%%%%%%%%%%%%%%%%%%%%%%%%%%%%%%%%%%%%%%%%%%%%%%%%%%%%%%%%%%%%%%%%%%%%%%%%%%%%%%%%%%%%%%%%%%%%%%%%%%%%%%%%%%%%%%%%%%%%%%%%%%%%%%

\subsection{Computational aspects}
\label{subsec:computation}

In this section we describe some technical details that need to be kept in mind 
when implementing our heuristics. Moreover, we explicitly derive the computational 
costs of our methods. 

There are several important points to keep in mind when implementing 
the methods described in the paper. %previous subsection.  
First of all, for the downdates, updates, and rewires 
based on the edge subgraph centrality we need to compute
the diagonal entries of $e^A$. This is the most expensive
part of these methods. There are, however, techniques that
can be used to rapidly estimate the diagonal entries of $e^A$ and
to quickly identify the top $\ell$ nodes, where $\ell \ll n$;
see \cite{A:Benzi2010,A:Benzi2013,A:Fenu2013} and references
therein. It should be pointed out that very high accuracy
is not required or warranted by the problem. 
We also recall that the same techniques (based on
quadrature rules and the Lanczos process) can be used
to compute the total communicability ${\bf 1}^T e^A {\bf 1}$
quickly (typically in $O(n)$ work), although such 
computation is actually not required
by any of the algorithms tested here except by the 
{\tt optimal} strategy, which is only used (for small
networks) as a baseline method. Such methods 
can also be used for rapidly estimating the node
total communicability centralities, $TC(i)= [e^A{\bf 1}]_i = {\bf e}_i^T e^A {\bf 1}$.

Secondly, when performing an update or the updating phase of 
a rewire, it makes sense to work with a subset of the set of all
virtual edges $\overline{E}$. Indeed, for a sparse network $\overline{E}$
contains $O(n^2)$ edges and for large $n$ this is prohibitive. 
Due to the particular selection criteria we want to use, it is
reasonable to restrict ourselves to the virtual edges in the subgraph of our network
that are incident to a subset $S$ of nodes containing
a certain percentage of the top nodes, ranked according to some
centrality measure. We found that for the larger networks
considered in the paper, using just the top $1\%$ 
of the nodes ranked using 
eigenvector centrality yields very good results.

Next, we derive the computational costs for the downdating techniques 
used in the accompanying paper.
Let $m$ be the number of edges in the network and let $K\ll n$, 
assumed bounded independently of $n$ as
$n\to \infty$, be the maximum number of modifications
we want to perform; in this paper, the maximum value of $K$
we consider is $2000$ (used for the three largest networks in our data set).

\begin{table}
\centering
\footnotesize
\caption{Computational costs for the downdating and updating techniques introduced in the accompanying paper.}
%[{\bf CUT?}--- $n$ and $m$ represent the number of nodes and  edges, 
%respectively. $K\ll n$ is the maximum number of modification 
%we want to perform and it is assumed to be bounded independently of $n$. 
%Update: $\ell=|S|$ is the number of nodes in the set $S\subset V$ considered.]}
\begin{tabular}{lcc}
\hline
Method                & Downdate      &   Update \\
\hline
{\tt optimal}         &  $O(Kn^2)$    &  $O(Kn^3)$\\
{\tt subgraph}        &  $O(Kn)$      &  $O(K\ell n)$\\
{\tt eigenvector}     &  $O(Kn)$      &  $O(K\ell n)$\\
{\tt nodeTC}          &  $O(Kn)$      &  $O(K\ell n)$\\
{\tt degree}          &  $O(Kn)$      &  $O(Kn)$\\
{\tt subgraph.no}     &  $O(n\log n)$ &  $O((K+\ell)n)$\\
{\tt eigenvector.no}  &  $O(n\log n)$ &  $O((K+\ell)n)$\\
{\tt nodeTC.no}       &  $O(n\log n)$ &  $O((K+\ell)n)$\\
\hline
\end{tabular}
\label{tab:costs}
\end{table}

In the {\tt optimal} method we remove each edge 
in turn, compute the total communicability after each downdate, and
then choose the downdate which caused the least decrease in $TC(A)$;
assuming that the cost of computing $TC(A)$ is $O(n)$, we find a total
cost of $O(Kmn)$ for $K$ updates. Since $m=O(n)$, this amounts to $O(Kn^2)$.

Next, we consider the cost of techniques based on subgraph centralities.
The cost of computing the node subgraph centralities is not easy to
assess in general, since it depends on network properties and on 
the approximation technique used. If a rank-$k$ approximation is used
\cite{A:Fenu2013}, the cost is approximately $O(kn)$; hence, the cost is
linear in $n$ if $k$ is independent of $n$, which is appropriate for
many types of networks. Computing the edge centralities requires another
$m=O(n)$ operations, and sorting the edges by their centralities costs
approximately $m\ln m$ comparisons. Note that sorting is only necessary
in the {\tt subgraph.no} variant of the algorithm; indeed, with
{\tt subgraph} we recompute the centralities after each update and
instead of sorting the result we only need to identify the edge
of minimum centrality at each step, which can be done in $O(m)$ work.  
Summarizing, the cost of {\tt subgraph} is $O(K(n+m)) = O(Kn)$ for 
$K$ downdates if we assume the subgraph centralities can be computed in
$O(n)$ time, and the cost for {\tt subgraph.no} is $O(n+m) = O(n)$ 
plus a pre-processing cost of $O(m\ln m)$ ($=O(n\ln n)$) 
comparisons for sorting
the edge centralities. Although the asymptotic cost of {\tt subgraph.no}
appears to be higher than that of {\tt subgraph} (due to the
$n\ln n$ term), in practice %%the prefactor
%%for the latter algorithm is so large 
one finds that {\tt subgraph.no} 
runs invariably much faster than {\tt subgraph} for all cases tested here.

The costs associated with {\tt eigenvector} and {\tt eigenvector.no}
scale like those of {\tt subgraph} and {\tt subgraph.no},
assuming that the dominant eigenvector ${\bf q}_1$ of a large
sparse $n\times n$ adjacency matrix can be approximated in $O(n)$ time.
For many real world networks this is a reasonable assumption, since in
practice we found that running a fixed number of Lanczos steps will give
a sufficiently good approximation of ${\bf q}_1$.
The prefactors can be expected to be much smaller for the methods based
on eigenvector centrality than for those based on subgraph centrality.

The costs for {\tt nodeTC} and {\tt nodeTC.no} are comparable to those
for {\tt eigenvector} and {\tt eigenvector.no}, with the same asymptotic
scalability.

Finally, the cost of {\tt degree} is $O(Km)$ and hence also $O(Kn)$ for
a sparse network. 

Note that the cost of checking that the connectivity is preserved
after each downdate does not affect these asymptotic estimates; indeed, 
using $A^*$ search \cite{A:HNR68} this can be done in $O(m)$ time and hence
the additional cost is only $O(n)$ for a sparse network.
Of course, if the removal of an edge is found to disconnect the network
we do not perform the downdate and move on to the next candidate edge.

We consider next the computational cost for the updating strategies. 
As before we let
$K$, assumed bounded independently of $n$ as
$n\to \infty$, be the maximum number of updates
we want to perform.

It can be easily shown that the {\tt optimal} method costs $O(Kn^3)$ operations.
To estimate the cost of the remaining methods,
we assume that the set $S\subset V$ consisting of the top $\ell = |S|$ 
nodes (ranked according to some centrality measure) is known. 
The cost of determining this set is asymptotically
dominated by the term $O(n\ln n)$, as we saw.
As already mentioned, $\ell$ will be equal to some fixed percentage
of the total number of nodes in the network.

Both {\tt subgraph} and {\tt eigenvector} cost $O(K\ell n)$ operations, 
provided a low rank approximation (of fixed rank) is used to estimate the
subgraph centralities. The same holds for {\tt nodeTC}.
Typically, the prefactor will be larger
for the former method. Since we assumed that $\ell = O(n)$ (albeit with
a very small prefactor, like $10^{-2}$) these methods exhibit
an $O(n^2)$ scaling.  In practice this is somewhat misleading, since the
quadratic scaling is not observed until $n$ is quite large. 

Finally, {\tt degree} costs $O(K\ell) = O(n)$ while
{\tt subgraph.no}, {\tt eigenvector.no} and
{\tt nodeTC.no} all cost $O((K+\ell)n)$.
Again, the latter cost is asymptotically quadratic but the actual
cost is dominated by the linear part until $n$ becomes quite large.
We note that we can obtain an asymptotically linear scaling by 
imposing an upper bound on $\ell$, i.e., on the fraction of nodes
that we are willing to include in the working subset $S$ of nodes.
We stress that because of the widely different prefactors for the various
methods, these asymptotic estimates should only be taken as 
roughly indicative. 
%In the next section we present timings showing the linear scaling behavior of the various
%heuristics in practice, at least for the networks considered here.

%%%%%%%%%%%%%%%%%%%%%%%%%%%%%%%%%%%%%%%%%%%%%%%%%%%%%%%%%%%%%%%%%%%%%%%%%%%%%%%%%%%%%%%%%%%%%%%%%%%%%%%%%%%%%%%%%%%%%%%%%%%
\subsection{Numerical tests: small networks}
In this section we present the results obtained when performing numerical tests on 
four networks of small size.  For these networks it is possible to apply the
{\tt optimal} strategy and to compare the other, more practical strategies
with it. The results of this comparison serve as a justification for
the use of our heuristics on larger networks. 

\begin{table}[t]
\footnotesize
\centering
\caption{Description of the Data Set.}
\label{A.tab:Datasets}
\begin{tabular}{cccccc}%{|c|c|c||c|c|c|}
\hline
NAME & $n$ & $m$ &  $\lambda_1$ & $\lambda_2$ & $\lambda_1-\lambda_2$ \\
\hline
Zachary & 34 & 78  & 6.726 & 4.977 & 1.749  \\
Sawmill & 36 & 62  & 4.972 & 3.271 & 1.701 \\
social3 & 32 & 80 & 5.971 & 3.810 & 2.161  \\
dolphins & 62 & 159 & 7.193 & 5.936 & 1.257 \\
\hline
\end{tabular}
\end{table}

The real-world networks used in the tests (see Table \ref{A.tab:Datasets}) 
come from a variety of sources. 
The Zachary Karate Club network is a 
classic example in social network analysis 
\cite{A:Zachary1977}. The Sawmill and social3 networks were provided to us 
by Prof.~Ernesto Estrada. 
The Sawmill network describes a communication network within a 
small enterprise (see \cite{A:Michael1997, A:Nooy2004}),
whereas social3 is a network of social contacts among college students 
participating in a leadership course (see \cite{A:Zeleny1950}).
The network dolphins (see \cite{A:Lusseau2003}) is in the Newman group from the Florida Sparse Matrix Collection \cite{A:Davis}  
and represents a social network of frequent associations between 62 dolphins in a 
community living in the waters off New Zealand.
Table \ref{A.tab:Datasets} reports the number of nodes ($n$), 
the number of edges ($m$), the 
two largest eigenvalues, and the spectral gap.
We use these networks to test all the greedy methods described
in the accompanying paper. %(except for {\tt optimal}, which is only
%applied to the four smallest networks) and the last three to illustrate the
%performance of the most efficient among the methods tested.

\begin{figure}[t]
\centering
\caption{Evolution of the normalized total communicability vs.~number 
of downdates performed on small networks.}
\label{fig:down}
\includegraphics[width=.9\textwidth]{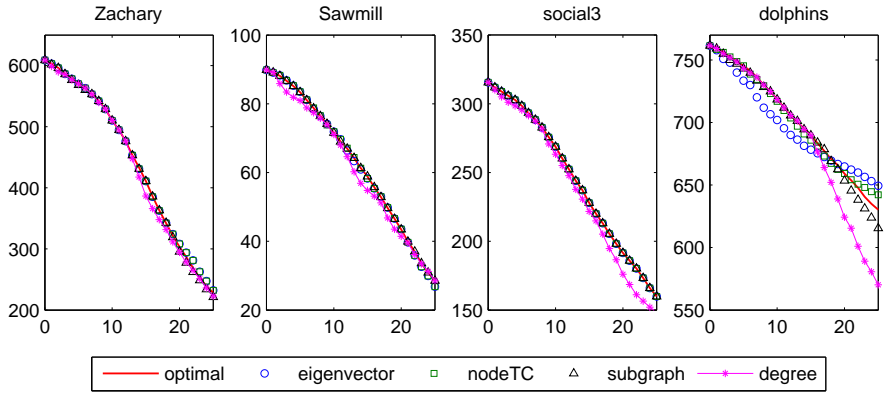}
\end{figure}

We begin by showing results for the four smallest networks.
Figure \ref{fig:down} displays the results
obtained with the downdating methods {\tt optimal}, {\tt eigenvector},
{\tt nodeTC}, {\tt subgraph}, and {\tt degree}. The results for {\tt eigenvector.no},
{\tt subgraph.no}, and {\tt nodeTC.no}
are virtually indistinguishable from those
obtained with {\tt eigenvector}, {\tt subgraph} and {\tt nodeTC} and are therefore
not shown.  At each step we modify the network by downdating an edge and we
then compute and plot the new value of the normalized total communicability.
The tests consist of 25 modifications.

%%The results for the downdating techniques are displayed in 
%%Figures \ref{fig:down} and \ref{fig:down_big}.
%%At each step we modify the network by downdating an edge and we 
%%then compute the new value of the normalized total communicability.
%%The tests consist of 25 modifications on the four smaller networks 
%%(Zachary, Sawmill, social3, and dolphins) and of 50 modifications on 
%%the medium size ones (Minnesota, USAir97,as-735 and Erd\"os02).

Figure \ref{fig:down} shows that our methods all perform similarly
and give results that are in most cases very close to those obtained
with {\tt optimal}, and occasionally even better,
as is the case for {\tt eigenvector}
(and {\tt eigenvector.no}) 
on the dolphins network after a sufficient
number of downdates have been performed. 
This result may seem puzzling at first, however, it can be easily 
explained by noticing that {\tt eigenvector} selects a different edge from 
that selected by {\tt optimal} at the third downdate step. 
Hence, from that point on the adjacency matrices on which the 
methods work are different, and the choice performed by the 
{\tt optimal} method may no longer be optimal for the 
graph manipulated by {\tt eigenvector}. Note that even the
simple heuristic {\tt degree} seems to perform well, except 
perhaps on the dolphins network after 15 or so downdate steps.
Overall, the methods based on eigenvector  
and total communicability centrality 
appear to perform best in view
of their efficacy and low cost. 

\begin{figure}[t]
\centering
\caption{Evolution of the normalized total communicability vs.~number of updates 
performed on small networks.}
\label{fig:up}
\includegraphics[width=.9\textwidth]{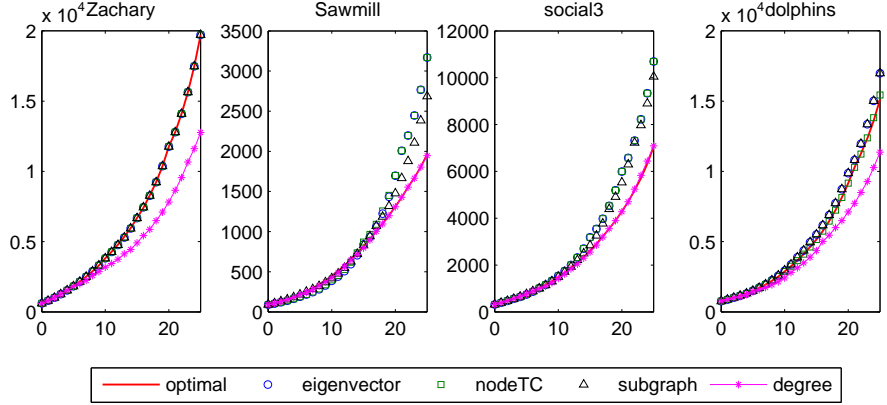}
\end{figure}

\begin{figure}[t]
\centering
\caption{Evolution of the normalized total communicability vs.~number 
of rewires performed on small networks.}
\label{fig:rew}
\includegraphics[width=.9\textwidth]{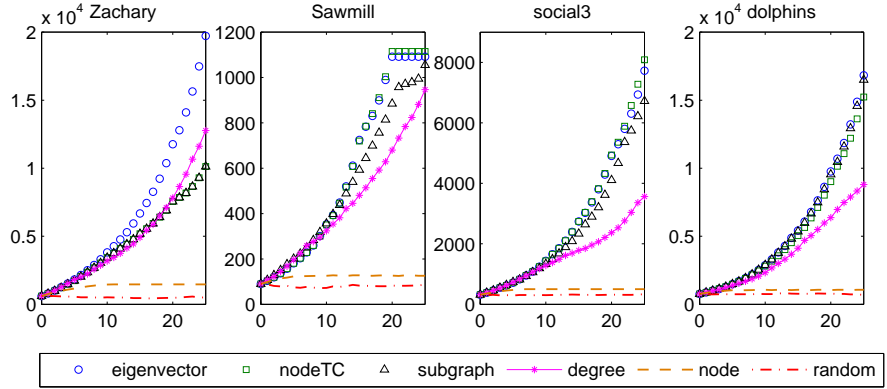}
\end{figure}

The results for the updating methods are reported in Figure \ref{fig:up}.
As for the downdating methods, {\tt subgraph.no}, {\tt eigenvector.no},
and {\tt nodeTC.no} 
return results that are virtually identical to those obtained using
{\tt subgraph}, {\tt eigenvector} and {\tt nodeTC}, therefore we omit them from the figure.
Once again we see that the methods based on eigenvector, subgraph, and total
communicability centrality 
give excellent results, whereas {\tt degree} is generally not as effective.

Rewiring results are displayed in Figure \ref{fig:rew}.
Clearly, the methods making use of edge centrality
perform quite well, in contrast to {\tt random}
rewiring (which is only included as a base for comparison).
Note also the poor performance of {\tt node}, showing that 
the use of edge centralities (as opposed to node centralities
alone) is indispensable in this context.

 The values obtained %%for the smaller networks using 
using the updates are in general higher than those obtained using the rewiring 
strategies, since updating implies the addition of edges whereas in
rewiring the number of edges remains the same. %The difference is
%especially pronounced for these small networks (except for dolphins), where 
For these methods 
the effects of downdates have a great impact,
leading to a decrease by up to nearly $70\%$ of the original value 
of the total communicability after 25 downdates (cf.~Figure \ref{fig:down}). 
It is noteworthy that the methods based on the edge eigenvector 
and total communicability
centrality appear to be more stable than the others 
under rewiring and to dampen the effect of the downdates even for small networks. 

%%%%%%%%%%%%%%%%%%%%%%%%%%%%%%%%%%%%%%%%%%%%%%%%%%%%%%%%%%%%%%%%%%%%%%%%%%%%%%%%%%%%%%%%%%%%%%%%%%%%%%%%%%%%%%%%%%%%%%%%%%%%%%%%%%
\subsection{Algorithms}
This section provides pseudocodes for all the algorithms used in the accompanying paper. 
Algorithms \ref{alg:down.check} and \ref{alg:down.no} implement our downdating techniques with or without the connectivity 
check, while 
Algorithm \ref{alg:up} implements our updating heuristics. 
All these three algorithms can be used with or without the recomputation of the rankings for the edges after 
each modification has been performed. 
They require as inputs the initial graph $G$ (typically in the form of
its adjacency matrix $A$) and a budget $K$, 
i.e., the number of modifications one 
wants to perform. 
The Boolean $greedy$ indicates whether the rankings of the edges have to be recomputed after each modification ($greedy=1$) or not ($greedy=0$). 

For the updating algorithm, it is also required to give in input 
a subset $S\subset V$ of nodes. 
The $K$ modifications will be selected among the virtual edges in the subgraph containing the nodes in $S$ and the corresponding edges. 

Finally, Algorithm \ref{alg:Chan} contains a detailed description of the technique introduced in \cite{A:Chan2014}. 
This algorithm requires as input the 
adjacency matrix $A$ and a budget $K$, as our methods do. 
Moreover, this methods requires as input an integer $t$, which is the number of leading eigenpairs 
to be considered and updated during the search for the $K$ modifications. %virtual edges to be added.

\begin{algorithm}[t]\label{alg:down.check}
\footnotesize
\caption{\footnotesize{Downdating algorithm with connectivity check.} }
%\SetLine 
% \SetAlgoLined
 \KwData{Initial graph $G$, $K\in\mathbb{N}$, $greedy\in\{0,1\}$}
 \KwResult{Set $\mathcal{S}$ of $K$ edges to be removed}
$\mathcal{S}=\emptyset$\; 
$c=0$\;
%$l=1$\;
$\mathcal{E}$ = list of edges in the graph\;

\eIf{greedy}{
  \While{$(c<K)\ \&\&\ (|\mathcal{E}|>0)$}
  { $found\_edge=0$\;
    Compute the centrality measure of interest $\forall (i,j)\in\mathcal{E}$\;
    \While{$(found\_edge == 0)\ \&\&\ (|\mathcal{E}|>0)$}{
      $G'=G$\;
      $s=$ element in $\mathcal{E}$ with the smallest centrality\;
      Downdate $s$ from $G'$\;
      \If{$G'$ is connected}{
        $G=G'$\;
        $ found\_edge = 1$\;
        $\mathcal{S} = \mathcal{S}\cup\{s\}$\;
        $c=c+1$\;
        }
      $\mathcal{E}=\mathcal{E}\setminus\{s\}$\;
      }
}
}{                 %case: greedy == 0
$l=1$\;
Compute edge centrality measure of interest $\forall (i,j)\in \mathcal{E}$\;
Sort the edges in ascending order\;
\While{$(c<K)\ \&\&\ (l\leq|\mathcal{E}|)$ }{
  %$found\_edge=0$\;
  %\While {$found\_edge==0$}{ 
    $G'=G$\;
    $s=$ $l$th edge in the sorted array\;
    Downdate $s$ from $G'$\;
    \If{$G'$ is connected}{
      $G=G'$\;
      %$found\_edge=1$\;
      $c=c+1$\;
      $\mathcal{S}=\mathcal{S}\cup\{s\}$\;}
    $l=l+1$\;}}%}
Return $\mathcal{S}$.
\end{algorithm}

\begin{algorithm}[t]\label{alg:down.no}
\footnotesize
\caption{\footnotesize{Downdating algorithm without connectivity check.} }
%\SetLine 
% \SetAlgoLined
 \KwData{Initial graph $G$, $K\in\mathbb{N}$, $greedy\in\{0,1\}$}
 \KwResult{Set $\mathcal{S}$ of $K$ edges to be removed}
$\mathcal{S}=\emptyset$\; 
$c=0$\;
%$l=1$\;
$\mathcal{E}$ = list of edges in the graph\;

\eIf{greedy}{
  \For{$iter=1:K$}{ 
    Compute the centrality measure of interest $\forall (i,j)\in\mathcal{E}$\;
    $s=$ element in $\mathcal{E}$ with the smallest centrality\;
    Downdate $s$ from $G$\;
    $\mathcal{S}=\mathcal{S}\cup\{s\}$\;
    $\mathcal{E}=\mathcal{E}\setminus\{s\}$\;
  }
}{                 %case: greedy == 0
Compute edge centrality measure of interest $\forall (i,j)\in \mathcal{E}$\;
Sort the edges in ascending order\;
$\mathcal{S}=$ top $K$ elements in the sorted array\;
}
Return $\mathcal{S}$.
\end{algorithm}

\begin{algorithm}[t]\label{alg:up}
\footnotesize
\caption{\footnotesize{Updating algorithm.} }
%\SetLine 
% \SetAlgoLined
 \KwData{Initial graph $G$, $K\in\mathbb{N}$, $S\subset V$ nodes in the subgraph, $greedy\in\{0,1\}$}
 \KwResult{Set $\mathcal{S}$ of $K$ edges to be added}
$\mathcal{S}=\emptyset$\; 
$\mathcal{E}$ = list of virtual edges in the subgraph containing nodes in $S$\;
\eIf{greedy}{
\For{iter = 1 : $K$}{
  Compute edge centrality measure of interest $\forall (i,j)\in\mathcal{E}$\;
  $s =$ element in $\mathcal{E}$ having the largest centrality\;
  Update $s$ in $G$\;
  $\mathcal{S}=\mathcal{S}\cup\{s\}$\;
  $\mathcal{E}=\mathcal{E}\setminus\{s\}$\;
  }}
{Compute edge centrality measure of interest $\forall (i,j)\in \mathcal{E}$\;
Sort the edges in descending order\;
$\mathcal{S}$= top $K$ elements in the sorted array\;
}
Return $\mathcal{S}$.
\end{algorithm}
\begin{algorithm}[t]
\footnotesize
\caption{\footnotesize{Updating algorithm from \cite{A:Chan2014}.} }
\label{alg:Chan}
%\SetLine 
% \SetAlgoLined
 \KwData{$A$ adjacency matrix, $K\in\mathbb{N}$, and $t\in\mathbb{N}$.}
 \KwResult{Set $\mathcal{S}$ of $K$ edges to be added}
$\mathcal{S}=\emptyset$\; 
Compute the top $t$ eigenpairs $(\lambda_k,\mathbf{q}_k)$ of $A$\;
\For{iter = 1 : $K$}{
  Compute $d_{\max}=\max(d_i)$, the largest row sum of $A$ \;
  Find the set $C$ of $d_{\max}$ nodes with the highest eigenvector centrality\;
  Select the edge $(i^*,j^*)\in\overline{E}$  that maximizes 
  $$
  e^{\lambda_1} \left( e^{2q_1(i)q_1(j)} + \sum_{h=2}^te^{\lambda_h-\lambda_1}e^{2q_h(i)q_h(j)}\right)
  $$
  and such that $i^*,j^*\in C$, $i^*\neq j^*$\;
  $\mathcal{S}=\mathcal{S}\cup \{(i^*,j^*)\}$, $E=E\cup \{(i^*,j^*)\}$\;
  Update $A$\;
  Update the top $t$ eigenpairs as
$$\left\{
\begin{array}{l}
\lambda_k=\lambda_k+2q_k(i)q_k(j); \\
\mathbf{q}_k=\mathbf{q}_k+\sum_{h\neq k}\left(\frac{q_h(i)q_k(j)-q_k(i)q_h(j)}{\lambda_k-\lambda_h}\mathbf{q}_h\right)
\end{array}\right.\qquad k=1,2,\ldots,t;$$
}
Return $\mathcal{S}$.
\end{algorithm}

\subsection{Free energy in networks}
In this section we recall the approach used in \cite{A:Estrada2007} to relate the 
Estrada index of a network with its Helmholtz free energy and, 
consequently, with its 
Gibbs entropy. 

Consider a network in which every edge is weighted by
a parameter $\beta> 0$ and consider its adjacency matrix $\beta A$.
The eigenvalues of this new matrix are $\beta\lambda_j$ for all
$j=1,2,\ldots,n$ and its Estrada index becomes $EE(G,\beta)=\text{Tr}(e^{\beta A})$,
where Tr denotes the trace.
This index can be interpreted as the {\it partition function} of the
corresponding complex network:
$$Z(G,\beta):=EE(G,\beta)=\text{Tr}(e^{\beta A}).$$
Form the standpoint of quantum statistical mechanics, 
$\mathcal{H}=-A$ is the system Hamiltonian and $\beta=\frac{1}{k_BT}$
is the inverse temperature, with $k_B$ the Boltzmann constant
and $T$ the absolute temperature.
It is well known \cite{A:EstradaBook,A:EHB12} that $\beta$ can be understood as a measure
of the ``strength'' of the interactions between pairs of vertices;
the higher the temperature (i.e., the lower the value of $\beta$),
the weaker the interactions. The eigenvalues $\lambda_i$ (for $i=1,\ldots,n$)
give the possible energy levels, each corresponding to
a different state of the system. 

The probability that the system is found in a particular state can
be obtained by considering the Maxwell--Boltzmann distribution:
$$p_i=\frac{e^{\beta\lambda_i}}{EE(G,\beta)},\quad i=1,\ldots ,n.$$
Using this notation and the fact that the Estrada index can be seen as
the partition function of the system, in \cite{A:Estrada2007} the authors
define the {\it Gibbs entropy} of the network as 
$$S(G,\beta)=-k_B\sum_{i=1}^np_i \ln(p_i)=-k_B\beta\sum_{i=1}^n(\lambda_i p_i)+
k_B\ln(EE(G,\beta))$$
where in the last equality we have used the fact that $\sum_ip_i=1$.

Using now the standard relation $F=H-TS$ that relates the {\it Helmholtz
free energy} $F$, the {\it total energy} of the network $H$, the Gibbs
entropy $S$, and the absolute temperature of the system $T$, the authors derive:
$$\left\{\begin{array}{l}
H(G,\beta)=-\sum_{i=1}^n\lambda_i p_i,\\
F(G,\beta)=-\beta^{-1} \ln(EE(G,\beta)).
\end{array}\right.$$
It is then clear that if we set $\beta=1$ and let $F:=F(G,1)$, we then get $F=-\ln(EE(G))$.

%\clearpage
%%%%%%%%%%%%%%%%%%%%%%%%%%%%%%%%%%%%%%%%%%%%%%%%%%%%%%%%%%%%%%%%%%%%%%%%%%%%%%%%%%%%%%%%%%%%%%%%%%%%%%%%%%%%%%%%%%%%%%%%%%%%%%%%%%%%%%%%%%%%%%%%%%%

\end{document}